\documentclass[a4paper]{amsart}

\usepackage{amssymb,amsmath,graphics,epsfig}
\usepackage{latexsym}
\usepackage{eucal}
\makeatletter
\@addtoreset{equation}{section}\makeatother

\newtheorem{theo}{Theorem}[section]
\newtheorem{lem}[theo]{Lemma}
\newtheorem{prop}[theo]{Proposition}

\newtheorem{defi}[theo]{Definition}

\newcommand{\norm}[1]{\left\Vert#1\right\Vert}
\newcommand{\beq}{\begin{equation}}
\newcommand{\eeq}{\end{equation}}
\newcommand{\bce}{\begin{center}}
\newcommand{\ece}{\end{center}}
\newcommand{\barr}{\begin{array}}
\newcommand{\earr}{\end{array}}
\newcommand{\ben}{\begin{enumerate}}
\newcommand{\een}{\end{enumerate}}
\newcommand{\li}{\mathcal{L}}
\newcommand{\rr}{\mathbb{R}}
\newcommand{\nn}{\mathbb{N}}
\newcommand{\cc}{\mathbb{C}}
\newcommand{\zz}{\mathbb{Z}}

\newcommand{\disp}[1]{\displaystyle{#1}}

\newcommand{\eco}{\varphi_{\alpha}}

\newcommand{\rd}{\mathbb{R}^{d}}
\newcommand{\rdd}{\mathbb{R}^{2d}}
\newcommand{\supp}{\mbox{Supp}}
\newcommand{\tr}{\mbox{Tr}}
\newcommand{\A}{\mathcal{A}}

\newcommand{\F}{\mathcal{F}}

\newcommand{\detp}{\mbox{det}^{-\frac{1}{2}}_+ }
\newcommand{\indic}{\operatorname{1\negthinspace l}}
\def\qed{\hfill$\square$}

\newcommand{\ud}{\frac{1}{2}}

\newcommand{\wt}{\widehat{W_t}}
\newcommand{\wtu}{\widehat{W_1}}
\newcommand{\wtd}{\widehat{W_2}}

\newcommand{\ope}{op\'erateur }

\newcommand{\grad}{\nabla H(z)}
\newcommand{\jgrad}{J\nabla H(z)}
\newcommand{\ome}{\Omega_0}
\newcommand{\omer}{\Omega_{red}}
\newcommand{\ce}{\mathcal{C}_{E}}

\newcommand{\nrjt}{\widetilde{\Sigma}_E}

\newcommand{\alg}{\mathcal{G}}
\newcommand{\carb}{ \overline{\chi (g)} }
\newcommand{\mg}{M(g)}
\newcommand{\mgm}{M(g^{-1})}
\newcommand{\gmu}{g^{-1}}
\newcommand{\la}{\lambda}
\newcommand{\e}{\varepsilon}
\newcommand{\aaa}{\alpha}
\newcommand{\gmm}{\gamma}

\newcommand{\nrj}{\Sigma_E}

\newcommand{\vfi}{\varphi}

\newcommand{\lde}{L^2(\mathbb{R}^d)}
\newcommand{\ldec}{L^2_{\chi}(\mathbb{R}^d)}

\newcommand{\h}{\mathcal{H}}

\newcommand{\op}{Op_h^w}
\newcommand{\mm}{\tilde{M}(g)}
\newcommand{\mmu}{\tilde{M}(g)^{-1}}
\newcommand{\hq}{\widehat{H}}
\newcommand{\hqr}{\widehat{H}_{\chi}}
\newcommand{\chg}{\Lambda_h}
\newcommand{\trp}{\mathcal{T}_h(\alpha)}

\newcommand{\gra}{\int_{\mathbb{R} ^{2d}_{\alpha}} }

\begin{document}
\title[Reduced Gutzwiller formula with symmetry: case of a Lie group]{Reduced Gutzwiller formula with symmetry: \\ case of a Lie group}
\author[Roch Cassanas]{Roch Cassanas}
\address{Laboratoire de Math\'ematiques Jean Leray\\
         Universit\'e de Nantes\\
         2, Rue de la Houssini\`ere, BP92208\\    
         F-44322, Nantes Cedex 03, France}
     \email{cassanas@math.univ-nantes.fr}
%
\begin{abstract}
\noindent 
We consider a classical Hamiltonian $H$ on $\rdd$, invariant by a Lie group of symmetry $G$, whose Weyl
quantization $\hq$ is a selfadjoint operator on $\lde$. If $\chi$ is an irreducible character of $G$, we
investigate the spectrum of its restriction $\hqr$ to the symmetry subspace $\ldec$ of $\lde$ coming from
the decomposition of Peter-Weyl.
We give semi-classical Weyl asymptotics for the eigenvalues counting function of $\hqr$ in an interval of $\rr$, and interpret
it geometrically in terms of dynamics in the reduced space $\rdd/G$.
Besides, oscillations of the spectral density of $\hqr$ are described by a Gutzwiller trace formula involving periodic
orbits of the reduced space, corresponding to quasi-periodic orbits of $\rdd$.
\end{abstract}
\maketitle

\section{Introduction}
The purpose of this work is to give semi-classical spectral asymptotics of a quantum Hamiltonian on $\rdd$
reduced by a compact Lie group of symmetry $G$. We will interpret coefficients in terms of reduced classical
dynamics in $\rdd/G$. This paper follows a preceding study on finite groups (see \cite{cras1},\cite{fini}). For a more
detailled introduction to the concepts, we refer the reader to this article.\

Mathematically, first systematic quantum investigations with symmetry reduction were carried out on 
a Riemannian compact manifold $M$ for the Laplacian, or for an elliptic differential operator, as it was done for a
compact Lie group of symmetry simultaneously by Donnelly (\cite{Don}) and Br\"uning  \& Heintze (\cite{B-H}) in
1978-79. They gave Weyl asymptotics of the eigenvalues counting function of the operator for high energy, and
interpreted the results in terms of the reduced space $M/G$.
The same study was done by Helffer and Robert in 1984-86 in $\rd$ for an elliptic pseudo-differential operator
with a finite or compact Lie group of symmetry (see \cite{He-Ro 2}, \cite{He-Ro 3}). The semi-classical version of this work was done by El
Houakmi and Helffer in 1984-91, still for Weyl asymptotics (see \cite{El.H}, \cite{El.H-He}). However the
computation of the leading term wasn't totally achieved, which forms one of the goals of this article.
Coming back to compact manifolds, at the end of the 80's, Guillemin and Uribe showed oscillations of the
spectral density of a reduced elliptic pseudo-differential operator could be described by a trace formula involving
periodic orbits of the reduced space (see \cite{G-U 1}, \cite{G-U 2}, \cite{G-U 3}).
Another aim of this paper is to give an analogue of this result in $\rd$ in the context of articles of
Helffer and Robert previously quoted, using a different method.
This investigation is also related to the work of Borthwick, Paul and Uribe \cite{B-P-U} (see also Charles
\cite{Ch}) with Toeplitz operator on K\"ahler manifolds: even if their point of view consists
in quantizing directly the Hamiltonian in the reduced space, following the theory of geometric quantization and symplectic
reduction of Kostant, Souriau, Guillemin et al, they give a reduced Gutzwiller formula in this context.\\\

We briefly recall our setting: $H:\rdd\to\rr$ is a smooth Hamiltonian.
$G$ is a compact Lie group of invertible linear applications of the configuration space $\rd$. It acts
symplectically on the phase space $\rd\times\rd$ by $M:G\to Sp(d,\rr)$ defined by:
\beq\label{action}
M(g)(x,\xi):=(g\, x,^tg^{-1}\xi)
\eeq
We assume that $G$ is a symmetry for $H$, i.e. $H$ is $G$-invariant:
\beq\label{Ginvariant}
H(M(g)z)=H(z), \qquad \forall g\in G, \quad \forall z\in\rdd.
\eeq
The Hamiltonian system associated to
$H$ is:
\beq\label{hamsyst}
\dot{z}_t=J\nabla H(z_t), \mbox{ where } J=\left(
\barr{cc}
0    & I_d\\
-I_d & 0
\earr\right).
\eeq
It has the property that its flow $\Phi_t$ commutes with the symmetry $\mg$ for all $g$ in $G$.\

From the quantum point of view, under suitable assumptions (see (\ref{hypCF})), the Weyl quantization of $H$
is given by (if $u\in\mathcal{S}(\rd)$):
\beq\label{pseudoweyl}
\op(H) u(x)=(2\pi h)^{-d}\int_{\rd}\int_{\rd}e^{\frac{i}{h}(x-y)\xi}
H\left(\frac{x+y}{2},\xi\right) u(y) dy d\xi,\quad \forall x\in\rd.
\eeq
In particular, $\op(H)$ is essentially selfadjoint on $\mathcal{S}(\rd)$ and we denote by $\hq$
its selfadjoint extension with domain $D(\hq)$.
$G$ acts on the quantum space $\lde$ through $\tilde{M}$ defined for $g\in G$ by  :
\beq\label{actionquantique}
\mm(f) (x):=f(\gmu x), \qquad \forall f\in\lde,\quad \forall x\in\rd.
\eeq
If $\chi$ is an irreducible character of $G$, we set $d_{\chi}:=\chi(Id)$. Then, the symmetry
subspace $\ldec$ associated to $\chi$ is defined as the image of $\lde$ by the projector:
\beq\label{projo}
P_{\chi}:=d_{\chi}\int_{G} \carb \mm dg,
\eeq
where $dg$ is the normalized Haar measure on $G$. Then $\lde$ splits into a Hilbertian sum of $\ldec$'s
(Peter-Weyl decomposition), and the property (\ref{Ginvariant}) implies that each
$\ldec$ is invariant by $\hq$. The restriction $\hqr$ of $\hq$ to $\ldec$, will be called here the
{\it reduced quantum Hamiltonian}.\

Our goal is to investigate the spectrum of $\hqr$ in a localized interval of energy as $h$ goes to zero.
Without symmetry, the celebrated `Correspondence Principle' roughly says that
semi-classical asymptotics link quantum objects defined through $\hq$ (as trace or eigenvalues counting
function) with quantities of the classical Hamiltonian system (\ref{hamsyst}) of $H$. In the framework of
symmetries, since specialists of classical dynamics are used to investigate (\ref{hamsyst}) in the quotient $\rdd/G$, we expect
that semi-classical asymptotics would link the spectrum of $\hqr$ with quantities of the Hamiltonian system in
$\rdd/G$, or more precisely in $\ome/G$, where $\ome\subset \rdd$ is the zero level of the momentum map of
$G$ (see section 2 for details).\

It is  easy to check that $\ome$ is invariant by the action of $G$. We denote the {\it reduced space} by:
\beq\label{omegared}
\omer:=\ome/G.
\eeq
Let $\pi:\ome\to \omer$ be the canonical projection on the quotient. Thanks to (\ref{Ginvariant}), we can
clearly define the reduced classical Hamiltonian $\widetilde{H}:\omer\to \rr$ by:
\beq\label{hamclassred}
\widetilde{H}(\pi(z)):=H(z), \quad \forall z\in \ome.
\eeq
The topological flow $\tilde{\Phi}_t:\omer\to\omer$ is defined for all $t$ such that the flow $\Phi_t$ of
$H$ exists, by:
\beq
\tilde{\Phi}_t(\pi(z)):=\pi(\Phi_t(z)), \quad \forall z\in \ome.
\eeq
Let $U$ be an open set of $\rdd$ invariant by the action of $G$. Under suitable assumptions
(mainly the fact that stabilizers are conjugate on $\ome\cap U$ --see Definition \ref{hyp}--), then $\ome\cap U$ is a smooth
submanifold of $\rdd$, $(\ome\cap U)/G\subset \omer$ inherits a structure of smooth symplectic
manifold from $\rdd$, such that the restriction of $\pi$ and $\widetilde{H}$ would be smooth, and such that
$\tilde{\Phi}_t$ would be the flow of the Hamiltonian $\widetilde{H}$. The Riemannian structure of $\ome$
also descends to the quotient,
and we get a notion of volume on $\omer$. Note that, since all stabilizers are conjugate on $\ome\cap U$,
all $G$-orbits of points of $\ome\cap U$ have the same dimension as submanifolds of $\rdd$. A first result is the following:
\begin{theo}\label{Comptage} Let $\e>0$, $E_1<E_2$ in $\rr$ and set $U:=H^{-1}(]E_1-\e,E_2+\e[)$.
Assume that conditions of symplectic reduction (definition \ref{hyp}) are fulfilled on $\ome\cap U$. Suppose that
$H^{-1}([E_1-\e,E_2+\e])$ is compact and that $E_1$ and $E_2$ are non critical values of 
$\widetilde{H}$.
Then, for small $h$'s, the spectrum of $\hqr$ is discrete in $I:=[E_1,E_2]$, and, if $N_{I,\chi}(h)$ denotes the number of
eigenvalues of $\hqr$ in $I$ (with multiplicity), then we have:
\beq\label{counting}
N_{I,\chi}(h)=(2\pi h)^{k_0-d} d_{\chi} Vol_{red}(\widetilde{H}^{-1}(I))
\left[{\rho_{\chi}}_{|_{H_0}}:\indic\right]
+ O(h^{k_0-d+1}).
\eeq
where $k_0$ is the common dimension of $G$- orbits on $\ome\cap U$, and $Vol_{red}$ is the Riemannian volume
on $\omer$.
The algebraic quantity $\left[{\rho_{\chi}}_{|_{H_0}}:\indic\right] $ is an integer described in Theorem
\ref{weak}.
\end{theo}
This result generalizes the ones of Donnelly and Br\"uning \& Heintze to the case of $\rd$ in a quantum semi-classical context.
It was already conjectured for high energies in \cite{He-Ro 3}, where authors gave theorical
asymptotics using BKW methods, as in \cite{El.H-He} for a semi-classical version.\\\

Our second main result is devoted to the oscillations of the spectral density of states of $\hqr$ in a
neighbourhood of an energy $E\in\rr$. We get a trace formula similar to the one of Guillemin and Uribe
(\cite{G-U 3}), involving a sum over periodic orbits with energy $E$ of the Hamiltonian system of
$\widetilde{H}$ in $\omer$.
We set:
\beq\label{objetsreduits}
\widetilde{\nrj}:=\{ \widetilde{H}=E\}\subset\omer \;\mbox{ and }\; \mathcal{L}_{red}(E):=\{t\in\rr:\exists
x\in\widetilde{\nrj}  :\tilde{\Phi}_t(x)=x.\}
\eeq
We suppose that:
\begin{itemize}
\item There exists $\delta E>0$ such that $H^{-1}([E-\delta E,E+\delta E])$ is compact.
\item $f:\rr\to\rr$ is such that its Fourier transform $\hat{f}$ is smooth and compactly supported.
\item $\psi:\rr\to\rr$ is smooth and compactly supported in $]E-\delta E,E+\delta E[$.
\end{itemize}
Then, for small $h$'s, $\psi(\hq_{\chi})$ is trace class and we denote by $\mathcal{G}_{\chi}(h)$ the well
defined following trace:
\beq\label{trass}
\mathcal{G}_{\chi}(h):=Tr\left(\psi(\hq_{\chi}) f\left( \frac{E-\hq_{\chi}}{h} \right) \right).
\eeq
If $t_0\neq 0$, let $\mathcal{P}_{red}(E,t_0)$ be the set of periodic orbits in $\widetilde{\nrj}$ admitting $t_0$ as a period.
\begin{theo}\label{Gutz}
Set $U:=H^{-1}(]E-\delta E,E+\delta E[)$. Suppose that hypotheses of symplectic reduction a fulfilled on
$\ome\cap U$. Moreover, suppose that periodic orbits of $\widetilde{\nrj}\subset\omer$ having a period in
$\supp\hat{f}$ are non degenerate (in the sense
of \cite{CRR}) for these periods. Suppose also that $0\notin\supp(\hat{f})$.\\
Then $\mathcal{G}_{\chi}(h)$ has a complete asymptotic expansion in powers of $h$ as $h\to 0^+$ (modulo oscillating terms of type 
$e^{\frac{i \aaa}{h}}$), whose coefficients are distributions in $\hat{f}$ with support in
$\mathcal{L}_{red}(E)\cap\supp\hat{f}$. Moreover, the first term is given by: $\mathcal{G}_{\chi}(h)=$
\beq
\psi(E)d_{\chi}
\sum_{{\tiny t_0\in\mathcal{L}_{\mbox{{\tiny red}}}(E)\cap\supp\hat{f} }} \hat{f}(t_0)
\sum_{{\scriptsize \bar{\gmm}\in \mathcal{P}_{red}(E,t_0)}} e^{\frac{i}{h}S_{\bar{\gmm}}(t_0)} 
\frac{1}{2\pi}
\int_{\Lambda_{\bar{\gmm},t_0}} \carb d(t_0,z,g) d\sigma_{\Lambda_{\bar{\gmm}}}(z,g)+ O(h).
\eeq
where $S_{\bar{\gmm}}(t_0):=\int_0^{t_0}p_t(z)\dot{q_t}(z) dt$ (with
$\Phi_t(z)=(q_t(z),p_t(z))\in\rd\times\rd$) doesn't
depend on $z$ such that $\pi(z)\in\bar{\gamma}$, 
and:
\beq
\Lambda_{\bar{\gmm},t_0}:=\{ (z,g)\in (\ome\cap\nrj)\times G \; :  \; \mg\Phi_{t_0}(z)=z \;\mbox{ and }\;
\pi(z)\in\bar{\gmm}  \}\subset\rdd\times G,
\eeq
The density $d(t_0,z,g)$ doesn't depend on $h$ and $\chi$ and is detailled in Theorem \ref{asymptheolie} using quantities describing the classical
dynamical system (\ref{hamsyst}).\\
\end{theo}

Note that, even if periodic orbits are non degenerate in the reduced space $\omer$, the ones of $\ome$ are
generally degenerate in $\rdd$. Indeed, elements of $G$ map a periodic orbit of $\ome$ into another one of
same period, which creates tubes of periodic orbits with same period and doesn't match with a non degenerate situation.\

The author expects that one can calculate the density $d(t_0,z,g)$ in terms of primitive period, Maslov
index, and of the energy restricted Poincar\'e map of the periodic orbit $\bar{\gmm}$.
This seems to be a non-trivial calculus we have been able to complete only for finite groups
for the moment (see \cite{fini}). See also the work of the physicist S.C. Creagh in \cite{Cr}.
If one omits the assumption of non-degeneracy, under hypothesis of 
`$G$-clean flow' (see Definition \ref{flop}), we still get an asymptotic expansion, which depends on the connected
components of the set:
\beq\label{critik}
\mathcal{C}_E=\{(t,z,g)\in \supp\,(\hat{f})\times \rdd\times G : z\in(\ome\cap \Sigma_E),
\mg \Phi_t(z)=z \}.
\eeq

As in the article on finite groups \cite{fini}, we will use the work of Combescure and Robert on coherent
states instead of a traditional WKB method. The structure of this paper is the following: In section 2, we
precise our setting to get a nice smooth structure on the reduced space $\omer$ and give examples of classical and quantum reduced
Hamiltonians. Section 3 is dedicated to the so called `weak asymptotics', i.e. the asymptotic expansion of
$\tr(f(\hqr))$ when $f(\hqr)$ is trace class. This will help us to compute geometrically the leading term of
(\ref{counting}) for Weyl asymptotics. Then we adapt the method of \cite{CRR} using coherent states,
which leads us to an application of the generalised stationary phase theorem in section 4. We find
optimal conditions (called $G$-clean flow conditions) to apply this theorem to our case and give theorical
asymptotics. In section 5, we describe the particular case where $\supp(\hat{f})$ is located near zero,
and get Theorem \ref{Comptage}. Section 6 is dedicated to the case where we suppose that periodic
orbits are non degenerate in the reduced space, and leads to Theorem \ref{Gutz}.\\\

\textbf{Aknowledgements}: We found strong motivation in the work of the physicist S.C. Creagh (\cite{Cr}).
I am deeply grateful to D. Robert for his help, comments and suggestions, and also thank G. Carron for the proof of a geometrical
lemma (Lemma \ref{gillescarron}).
\section{Symplectic and quantum reduction}
\subsection{Symplectic reduction} Let $H:\rdd\to\rr$ be a smooth Hamiltonian invariant by a compact Lie group
$G$ of dimension $p$ as in section 1. The theorem of E.Noether says that, if $G$ is not finite, then $G$ provides the system
(\ref{hamsyst}) with integrals. In our case, these integrals are easy to compute. If $A\in \alg$,
the Lie algebra of $G$, we define $M(A)$ as:
\beq
M(A):=\left(
\barr{cc}
A & 0\\
0 & -^tA
\earr\right).
\eeq
Then, if $F_A:\rdd\to\rr$ is defined by $F_A(z):=\ud <JM(A)z,z>$, then $F_A$ is a first integral of $H$. Indeed,
one can differentiate at $t=0$ the identity:
$$H(e^{tM(A)}z)=H(z),$$
to get $\{H,F_A \}=0$, where $\{,\}$ denotes the Poisson bracket ($JM(A)$ is symmetric since for $g\in G$,
$M(g)$ is symplectic). Now, if $A_1,\dots,A_p$ is a basis of $\alg$, then we define the {\it momentum map}
$\mathbb{F}:\rdd\to\rr^p$ by $\mathbb{F}:=(F_{A_1},\dots,F_{A_p}).$
The zero level of $\mathbb{F}$ is:
\beq\label{ome}
\Omega_0:=\mathbb{F}^{-1}(\{ 0 \})=\underset{A\in\alg}{\bigcap}F_A^{-1}(\{ 0 \}).
\eeq
One can remark that $\ome$ is homogeneous and that $0\in\ome$. Furthermore, since $M$ is symplectic, it is
easy to check that $\ome$ is invariant by the action of $G$. Hence we can define $\omer$ as in
(\ref{omegared}), and $\pi$ as the projection on this quotient. We want to get, roughly speaking, a smooth
structure on $\omer$.
A natural hypothesis would be to suppose that $\ome$ is a manifold and that all stabilizers are conjugate
on $\ome$ (see for example \cite{K}, Theorem 4.18 p.196). Nevertheless, as we already said, zero is in $\ome$, and
its stabilizer is $G$ itself since the action is linear. One can also make this hypothesis only
on $\ome\setminus\{ 0 \}$. However, this would exclude the case of a cylindrical symmetry as we will see below. A less
restrictive hypothesis consists in demanding a smooth structure only on  a part of $\omer$, which will be
enough for our quantum application.
\begin{defi}\label{hyp}
If $U$ is an open set of $\rdd$ invariant by the action of $G$, we say that `hypotheses
of reduction are satisfied on $U\cap\ome$' if $U\cap \ome\neq \emptyset$ and if
following assumptions are fulfilled:
\begin{itemize}
\item Stabilizers of points of $\ome\cap U$ are all conjugate subgroups of $G$.
\item $\forall z\in \ome\cap U$, $\dim(\ome\cap U)=2d-\dim(G(z))$.
\end{itemize}
\end{defi}
\noindent We then have the following theorem generalizing the case of a free action originally given by J.Marsden and
A.Weinstein (one can find a proof of this theorem in the book \cite{Or-Ra}, Theorem 8.1.1 p.202):
\begin{theo}\label{symplecticred}(Symplectic reduction)\\
If $U$ is an open set of $\rdd$ such that hypotheses of reduction are
satisfied on $U\cap\ome$, then $U\cap \ome$ is a smooth submanifold of $\rdd$, and there exists a unique
structure of smooth manifold on $(\ome\cap U)/G\subset \omer$ such that the restriction of $\pi$ to $\ome\cap
U$ is a smooth submersion. Moreover, there exists a unique symplectic form $w_{red}$ on $(\ome\cap U)/G$ such that
$\pi^*w_{red}$ is the restriction of $<J.;.>_{\rdd}$ to $\ome\cap U$. Finally, the restriction  $\widetilde{H}:(\ome\cap
U)/G\to \rr$ is smooth, $\pi$ maps the integral curves of (\ref{hamsyst}) lying in $\ome\cap U$ on those of
the Hamiltonian system induced by $\widetilde{H}$ on $(\omer,w_{red})$, and $\tilde{\Phi}_t$ is the flow of
$\widetilde{H}$.
\end{theo}
Under hypotheses of definition \ref{hyp}, we will denote by:
\beq
H_0\subset G \mbox{ the stabilizer of one point $z_0$ in } \ome\cap U, \mbox{ and } k_0:=\dim(G(z_0)),
\eeq
$k_0$ being then the common dimension of $G$-orbits in $\ome\cap U$. Thus we have $\dim((\ome\cap
U)/G)=2(d-k_0)$. Note that $(\ome\cap U)/G$ inherits of a Riemannian structure from the submersion
$\pi:\ome\cap U\to (\ome\cap U)/G$, which in particular gives a measure on this set. Moreover, by a
dimensional argument, we have $T_z\ome=(J\alg z )^{\perp}$.\

\textsl{Remark}: most of time we will take $U:=H^{-1}(I)$, where $H$ is $G$-invariant and $I$ is an open
interval of $\rr$. Thus, hypotheses of reduction won't be fulfilled if $0\in H^{-1}(I)$. This would
require an additional treatment as it was done in \cite{El.H-He}.\\\\
Now, we give examples of classical reduction:\\
--\underline{{\it Spherical symmetry}}: $G=SO(d)$, $p=\frac{d(d-1)}{2}$. It is easily seen that we have the following
results:
$$\ome=\{ (x,\xi)\in\rd\times\rd : \mbox{ vectors $x$ and $\xi$ are linearly dependant } \}.$$
Stabilizers are conjugate in $\ome\setminus \{ 0 \}$, which is a submanifold of dimension $d+1$ of $\rdd$,
and thus we have $\dim(\ome\setminus \{ 0 \})=2d-\dim(G(z))$ for all $z$ in $\ome\setminus \{ 0 \}$.
Hence, for an invariant open set $U$ in $\rr^d\times\rr^d$ such that $\ome\cap U\neq\emptyset$, hypotheses
of reduction are fulfilled on $\ome\cap U$ if and only if $0\notin U$.\\
--\underline{{\it Cylindrical symmetry}}: $d=3$ and $G$
is the group of rotations of $\rr^3$ around the vertical axis. It is also easy to see that:
$$\ome=\{(x,\xi)\in\rr^3\times\rr^3 \, : \, (x_1,x_2)\mbox{ and }(\xi_1,\xi_2) \mbox{ are colinear }\}.$$
$\ome$ is the disjoint union of two sets $\Omega_1$ and $\Omega_2$ with:
$$\left\{ \barr{ll}
\Omega_1:=\{ (0,0,x_3;0,0,\xi_3)\, :\, x_3\in\rr,\; \xi_3\in\rr \}.\\
\Omega_2:=\{ (x^{\prime},x_3;\xi^{\prime},\xi_3) \, : \, x^{\prime} \mbox{ and } \xi^{\prime} \mbox{ are colinear in $\rr^2$
, and } (x^{\prime},\xi^{\prime})\neq 0 \}.
\earr\right.$$
On $\Omega_1$, stabilizers are conjugate to $G$ itself, whereas on $\Omega_2$  they are equal to
$\{Id_{\rr^3}\}$. Moreover, $\Omega_1$ is a plane and $\dim(\Omega_2)=5=2d-G(z)$ if $z\in\Omega_2$. Thus for
an invariant open set $U$ in $\rr^3\times\rr^3$ such that $\ome\cap U\neq\emptyset$, hypotheses of reduction
are fulfilled on $\ome\cap U$ if and only if $U\cap\Omega_1=\emptyset$.
\subsection{Quantum reduction} 
For general background on the decomposition of Peter-Weyl, reduced
Hamiltonians and interpretation of symmetry, we refer the reader to \cite{fini}, \cite{these}, \cite{He-Ro
3}, and \cite{Si}, \cite{Pi}). We just recall that if $\widehat{G}$ denotes the set of irreducible
characters
of $G$, then $G$ is countable (since $G$ is compact), and with the definition of $\ldec$ given in section
$1$, we have the Hilbertian decomposition:
\beq
L^2(\rr^d)=   \bigoplus_{ \mbox{ {\scriptsize $\chi\in\widehat{G}$ } }}^{\perp}  \quad \ldec,
\eeq
which comes from the identity:
\beq\label{metaplec}
\mmu \op(H)\, \mm=\op(H\circ \mg), \quad \forall g\in G.
\eeq
Basic properties of the $\ldec$'s are the same as in the case of finite groups, excepted for the spectrum
inclusion $\underset{\chi\in\hat{G}}{\cup }\sigma(\hqr)\subset \sigma(\hq)$,  without equality in general
(see \cite{these} p.11-12).\

Then we give examples of quantum reduction: for a character $\chi$ of
degree $1$, we have a simple description of $\ldec$ (see \cite{He-Ro 3} or \cite{fini}):
$$\ldec=\{ f\in \lde \;: \; \mm f=\chi(g) f \}.$$
This is in particular the case with $\chi=\chi_0$, the character of the trivial representation of $G$, i.e.
the representation of degree $1$, constant equal to identity. If we endow $\rd/G$ with the image measure of
the Lebesgue measure on $\rd$ by the canonical projection $\pi$ on the quotient, we note that the map $u\mapsto u\circ
\pi$ identify $L^2_{\chi_0}(\rd)$ with $L^2(\rd/G)$.
When the group is abelian, characters are of degree $1$ (see \cite{Se}). This is the case for $G=SO(2)$ with
$d=2$. One can show that $\widehat{G}$ is indexed by $\zz$, and if $R_\theta$ is the rotation of angle $\theta$ in
the anticlockwise sense, then $\widehat{G}=(\chi_n)_{n\in\zz}$ with $\chi_n(R_\theta):=e^{in\theta}$ and:
$$L^2_{\chi_n}(\rr^2)=\{ f\in L^2(\rr^2):\tilde{f}(r,\theta)=e^{-i n\theta}\, g(r),\; g\in
L^2(\rr_+,rdr) \},$$
where $\tilde{f}$ is the expression of $f$ in polar coordinates.
Using the expression of the Laplacian in spherical coordinates, if $H(x,\xi):=|\xi|^2+V(x)$ where
$V(x)=V_0(|x|)$ is radial in $\rr^2$, then (with $h=1$) we get that $\hq_{\chi_n}$ is unitary equivalent to the
following operator on $L^2(\rr_+,r dr)$:
$$ -\partial_r^2-\frac{1}{r}\partial_r+\frac{n^2}{r^2}+V_0(r).$$
An example with characters of arbitrary high degrees is  given by $G=SO(3)$. Conjugation classes are given by
rotations with same angle (non oriented). Let $\tilde{R}_\theta$ denotes the rotation with angle $\theta$ around
the vertical axis, then  $\widehat{G}=(\chi_n)_{n\in \nn}$ with:
$$\chi_n(\tilde{R}_{\theta}):= \sum_{k=-n}^n e^{i k\theta}=
\sin( (2n+1) \theta/2 )\, / \, \sin(\theta/2),
 \quad\theta\in [0,2\pi].$$
Hence, $d_{\chi_n}=2n+1$. Moreover, we can describe the symmetry subspaces using spherical harmonics $(Y_{n,k})$,
with $n\in\nn$ and $k\in\{-n,\dots,n\}$ ($n$ is  the quantum azimuthal number), which are the eigenfunctions
of the Laplacian in spherical coordinates ($-\Delta_{\mbox{sph}}Y_{n,k}=n(n+1)Y_{n,k}$):
$$ L^2_{\chi_n}(\rr^3)=\{ f\in L^2(\rr^3):\tilde{f}(r,\theta,\vfi)=g(r)\sum_{k=-n}^n \la_k\,
Y_{n,k}(\theta,\vfi),\quad \la_k \in \cc,\; g\in L^2(\rr_+,r^2dr)\}.$$
where $\tilde{f}$ is the expression of $f$ in spherical coordinates (see \cite{these} p.7-8 and \cite{Pi}
p.115). If $H(x,\xi):=|\xi|^2+V(x)$ where $V(x)=V_0(|x|)$ is radial in $\rr^3$, then (with $h=1$) we get that
$\hq_{\chi_n}$ is unitary equivalent to the following operator on $L^2(\rr_+,r^2dr)$:
$$ -\partial_r^2-\frac{2}{r}\, \partial_r+\frac{n(n+1)}{r^2} + V_0(r).$$
\section{Weak asymptotics}\label{faible}
The following result is interesting in itself and is also a usual way to compute the first term of the
asymptotic expansion of the eigenvalues counting function of $\hqr$ (Theorem \ref{Comptage}).
\begin{theo}\label{weak}
Let $G$ be a compact Lie group of $Gl(d,\rr)$ and  $H:\rdd\to \rr$ be a smooth Hamiltonian $G$-invariant
satisfying (\ref{hypCF}).
Let $E_1<E_2$ be such that $H^{-1}([E_1-\e,E_2+\e])$ is compact (where $\e>0$).
If $U:=H^{-1}(]E_1-\e,E_2+\e[)$ is such that hypotheses of reduction are satisfied on $\ome\cap U$,
if $f:\rr\to\rr$ is smooth, compactly supported in $]E_1,E_2[$, if $\chi\in\hat{G}$, then, for small $h$'s,
$f(\hqr)$ is trace class and $\tr(f(\hqr))$ has a complete asymptotic expansion in powers of
$h$ as $h\to 0^+$, with first term:
\beq
\tr(f(\hqr))=(2\pi h)^{k_0-d}d_{\chi} \int_{\omer}f(\widetilde{H}(x))
d\sigma_{red}(x). \left[{\rho_{\chi}}_{|_{H_0}}:\indic\right] \quad + \;O(h^{k_0-d+1}).
\eeq
Here $d\sigma_{red}$ is the measure corresponding to the Riemannian structure on $(\ome\cap U)/G$,
$\widetilde{H}$ is given by (\ref{hamclassred}), $k_0$ is the
common dimension of $G$-orbits on $\ome\cap U$, and $\left[{\rho_{\chi}}_{|_{H_0}}:\indic\right] $ is an
integer. Namely, if $H_0$ is any stabilizer of $\ome\cap U$ and $\rho_{\chi}$ is any representation with
character $\chi$, then  it is the number of times that the trivial representation $\indic$ is contained in
the decomposition into irreducible representations of $\rho_{\chi}$ restricted to $H_0$.
\end{theo}
Next sections are dedicated to the proof of this theorem.
\subsection{First reductions}
Suppose that $H$ satisfies the following assumptions (where $C,\, C_\aaa>0$),
\beq\label{hypCF}
\left\{
\begin{array}{l}
<H(z)>\leq C<H(z^{\prime})>.<z-z^{\prime}>^m, \quad \forall z,z^{\prime} \in \rr^{2d}.\\
|\partial_z^{\alpha} H(z)|\leq C_{\alpha}<H(z)>, \quad\forall z \in \rr^{2d}, \forall\alpha \in
\nn^{2d}.\\
H \mbox{ has a lower bound on }\rdd.
\end{array}
\right.
\eeq
where $<z>:=(1+|z|^2)^{\ud}$. Then, since $\supp(f)\subset ]E_1,E_2[$, we can write for $N_0\in\nn$ (see \cite{He-Ro 1}):
\beq\label{hero2}
f(\hq)= \sum_{j=0}^{N_0} h^j \op(a_j) + h^{N_0+1}\, R_{N_0+1}(h).
\eeq
where $\supp (a_j) \subset H^{-1}(]E_1-\e,E_2+\e[)$, $a_0(z)=f(H(z))$, with
$\disp{ \underset{0<h\leq 1}{ \mbox{Sup} } \norm{R_{N_0+1}(h)}_{\tr} \leq C\, h^{-d} }.$\
We easily get that all $a_j$ are $G$-invariant. Thus, we are led to show that, if $a$ is a smooth $G$-invariant function compactly supported in $U\subset \rdd$, and
$\widehat{A}:=\op(a)$, then $\tr(\widehat{A}_{\chi})$ has a complete asymptotic expansion in powers of $h$ when
$h\to 0^+$, and we have:
\beq
\tr(\widehat{A}_{\chi})=(2\pi h)^{k_0-d}d_{\chi} \int_{\omer}\widetilde{a}(x)
d\sigma_{red}(x). \left[{\rho_{\chi}}_{|_{H_0}}:\indic\right] \quad + O(h^{k_0-d+1}).
\eeq
where $\widetilde{a}(\pi(z)):=a(z)$.
By the trace formula using coherent states (see \cite{fini}, \cite{CRR}), we have:
$$\tr(\widehat{A}_{\chi})=\tr(\widehat{A}\,P_\chi)=d_\chi(2\pi h)^{-d}\int_G \int_{\rdd}\carb <\widehat{A}\,\eco,\mmu \eco> d\aaa dg.$$
As in \cite{CRR}, we can show that there exists a compact set $K$ in $\rdd$ such that:
$$\int_{\rdd\setminus K}|<\widehat{A}\,\eco;\mmu\eco>_{\lde}| d\aaa= O (h^{\infty}),$$
uniformly in $g\in G$. In view of Lemma 3.1. of \cite{CRR}, if $N\in\nn^*$, then there exists $C_{d,N}>0$
with:
$$\norm{ \op (a)\eco -\sum_{k=0}^N h^{\frac{k}{2}}\sum_{\gmm \in \nn^{2d}, |\gmm|=k}
\frac{ \partial^{\gmm}a(\aaa) }{\gmm !}\, \Psi_{\gmm,\aaa} }_{\lde} \leq
C_{d,N}.h^{\frac{N+1}{2}}.$$
where $\Psi_{\gmm,\aaa}:=\trp\chg Op_1^w (z^{\gmm})\tilde{\psi}_0$. For notations on coherent states, we
refer to \cite{CRR} and \cite{fini}. We can suppose that $\supp(a)\subset K$ and if $\chi_1$ is a smooth
function compactly supported in $\rdd$ with $\chi_1=1$ on $K$, writing $1=\chi_1+(1-\chi_1)$, we get:
$$\tr(\widehat{A}_{\chi})=d_\chi(2\pi h)^{-d}\sum_{k=0}^N h^{\frac{k}{2}}
\sum_{\gmm \in \nn^{2d}, |\gmm|=k}\int_G \int_{\rdd}\carb
\frac{\partial^{\gmm}a(\aaa)}{\gmm !} m_{\gmm}(\aaa,g) d\aaa dg\; + \; O(h^{\frac{N+1}{2}-d}),$$
where
$$m_{\gmm}(\aaa,g):=<\trp\chg Op_1^w(z^{\gmm})\tilde{\psi}_0;\mmu\trp\chg \tilde{\psi}_0>.$$
Thanks to the metaplectic property of $\tilde{M}$ -- see (\ref{metaplec}) --, since
$\trp=\op(\exp(\frac{i}{h}(px-q\xi))$ (if $\aaa=:(q,p)$), we can write:
$$\mmu \mathcal{T}_h(\aaa)=\mathcal{T}_h(\mgm \aaa) \mmu.$$
Let $Q_\gmm$ be the polynomial such that $Op_1^w(z^{\gmm})\tilde{\psi}_0=Q_{\gmm}\tilde{\psi}_0$. By using
formulas on coherent states of \cite{fini} or \cite{these}, we get:
\beq\label{mgamma}
m_{\gmm}(\aaa,g)=e^{-\frac{i}{2h}<J\mgm\aaa,\aaa>}
<\mathcal{T}_1\left(\frac{\aaa-\mgm\aaa}{\sqrt{h}}\right) Q_{\gmm}\tilde{\psi}_0;\mmu\tilde{\psi}_0 >.
\eeq
Here, we can use the same trick as in \cite{fini} which allows us to suppose that $G$ is composed of
{\it isometries} (we recall that, since $G$ is compact, by an averaging argument, it is conjugate to a subgroup
of the orthogonal group, and one is led to consider isometries by making a change of Hamiltonian). By a
clear change of variables in the integral given by the scalar product in (\ref{mgamma}), we get:
$$m_{\gmm}(\aaa,g)=(\pi h)^{-\frac{d}{2}} e^{-\frac{i}{2h}<J\mgm\aaa,\aaa>}
e^{\frac{i}{2h}<(I-^tg)p,(I-\gmu)q>} e^{ -\frac{1}{2h} |(I-\gmu)q|^2}$$
$$\times
\int_{\rd} e^{-\frac{|y|^2}{h}} e^{\frac{i}{h}<i(I-\gmu)q+(I-^tg)p;y>}
Q_{\gmm}\left(\frac{y}{\sqrt{h}}\right) dy.$$
Note that, since $Q_\gmm$ has the same parity as $|\gmm|=k$, only {\it entire} powers of $h$ will arise in
the asymptotics. Then, after making the change of variable $y^{\prime}=\frac{y}{\sqrt{h}}$, we write:
$$Q_{\gmm}(x)=\sum_{|\mu|\leq |\gmm|}\kappa_{\mu,\gmm} x^{\mu}.$$
We set $\beta_0:=i[i(I-\gmu)q+(I-^tg)p]$ and use the calculus of the Gaussian given by Lemma 3.2 of
\cite{fini} to get:
$$m_{\gmm}(\aaa,g)=e^{\frac{i}{h}\Phi(\aaa,g)} \sum_{|\mu|\leq |\gmm|}\kappa_{\mu,\gmm} \sum_{\eta\leq \mu} h^{-\frac{\eta}{2}}
(\ud\beta_0)^{\eta} P_{\eta}(2I_d),$$
where $P_{\eta}$ is a polynomial independant of $h$, $\aaa$ and $g$, with $P_0(2I_d)=1$, and:
\beq
\Phi(\aaa,g):=<B\aaa,\aaa> \;\mbox{ with } \; B=\frac{1}{4}J(\mg-\mgm) +\frac{i}{4}(I-\mg)(I-\mgm).
\eeq
Finally, we have:
\beq\label{viso}
\tr(\widehat{A}_{\chi})=
(2\pi h)^{-d}d_{\chi}\sum_{k=0}^N h^{\frac{k}{2}}\sum_{\gmm \in \nn^{2d}, |\gmm|=k}
\sum_{|\mu|\leq |\gmm|}\sum_{\eta\leq \mu} \frac{\kappa_{\mu,\gmm}}{\gmm !}P_{\eta}(2I_d)
h^{-\frac{|\eta|}{2}}I_{\gmm, \eta}(h) + O(h^{\frac{N+1}{2}-d}).
\eeq
with :
\beq\label{viso2}
I_{\gmm, \eta}(h):=\int_G \int_{\rdd}e^{\frac{i}{h}\Phi(\aaa,g)}\carb\partial^{\gmm}a(\aaa) F(\aaa,g)^{\eta}d\aaa dg
\eeq
\beq
F(\aaa,g):=\frac{i}{2}[i(I-\gmu)q +(I-^tg)p].
\eeq
Hence, we are led to a stationary phase problem, to find an asymptotic expansion of $I_{\gmm, \eta}(h)$. Note
that we will see that $F(\aaa,g)=0$ on the critical set of the phase. Thus, the asymptotics of $I_{\gmm,
\eta}(h)$ will start with a shift of $h^{\frac{|\eta|}{2}}$, which will compensate the term
$h^{-\frac{|\eta|}{2}}$ in (\ref{viso}).
\subsection{Phase analysis}
We want to apply the generalised stationary phase theorem in the form of \cite{CRR}. First we note that, if
$(z,g)\in\rdd\times G$, then $\Im\Phi(z,g)=\frac{1}{4}|(I-\mgm)z|^2 \geq 0$. In the rest of the article, we
often won't make the difference between $A\in\alg$ and $M(A)$ (or between $g\in G$ and $M(g)$) in order to
lighten notations.\\\\
-- {\it The critical set}: it is easily seen that 
$$\Im \Phi(z,g)=0\iff\partial_{z}\Phi(z,g)=0 \iff\mg z=z.$$
Moreover, if $A\in\alg$, and $g\in G$, then:
$$4\partial_g\Phi(z,g)(Ag)=<[J(Ag+\gmu A) -iAg(I-\gmu) + i(I-g)\gmu A]z,z>.$$
If, besides, $\Im \Phi(z,g)=0$, then: $\partial_g\Phi(z,g)(Ag)=0\iff <JM(A)z;z> =0$.
Thus we have proved that the critical set $\Gamma_0:=\{\Im \Phi=0\}\cap \{ \nabla \Phi=0\}\cap \{ U\times G
\}$\footnote{By the non stationary phase theorem we can restrict ourselves to $(z,g)\in\rdd\times G$ such that $z\in U$,
since we have from (\ref{hero2}): $supp(a)\subset U=H^{-1}(]E_1-\e,E_2+\e[)$.} satisfies:
\beq\label{Gamma_0}
\Gamma_0=\{ (z,g)\in \ome\times G : \mg z=z \mbox{ and } z\in U \}.
\eeq
\begin{lem}\label{varietefixe}
$\Gamma_0$ is a smooth submanifold of $\rdd\times G$ of dimension $\dim \Gamma_0=2d-2k_0+p$, where $p$ is the
dimension of $G$. Moreover, if $(z,g)\in\Gamma_0$, then:
\beq
T_{(z,g)}\Gamma_0=\{ (\aaa,Ag):\aaa\in T_z\ome,  A\in\alg \mbox{ and } (\mg-I)\aaa + Az=0 \}.
\eeq
\end{lem}
\underline{{\it Proof}} : we will make deep use of the reduction theorem \ref{symplecticred}.
Set:
$$R_0:=\{(z,\mg z):z\in\ome\cap U, \; g\in G\}.$$
We note that $R_0=(\pi\times\pi)^{-1}(diag [(\ome\cap U)/ G])$. By Theorem \ref{symplecticred},
$\pi\times\pi$ is a submersion on $(\ome\cap U)^2$, thus $R_0$ is a submanifold of dimension $\dim R_0=2d$,
and if $\mg z=z\in\ome\cap U$, then we have:
\beq\label{tangentgamma}
T_{(z,z)}R_0=\{ (\aaa,\beta)\in T_z\ome\times T_z\ome : \, d_z\pi(\aaa)=d_z\pi(\beta) \}.
\eeq
From the fact that $\pi$ is a submersion, we also deduce that $\forall z\in (\ome\cap U), \; \ker
d_z\pi=\alg z$. Moreover, one can differentiate the identity $\pi(\mg x)=\pi(x)$ with respect to $x\in\ome$ to get that,
if $\mg z=z$, then $\forall \aaa\in T_z\ome, \; (\mg-I)\aaa\in \alg z$. Thus, by (\ref{tangentgamma}) we get:
\beq\label{tgtR0}
T_{(z,z)}R_0=\{(\aaa,\mg \aaa+Az):\aaa\in T_z\ome, A\in\alg \}.
\eeq
Let $\vfi_0 :\left\{
\barr{ll}
\ome\times G \to R_{0}\\
(z,g)\mapsto (z,\mg z)
\earr\right.$.
We have: $diag(\ome\cap U)\subset R_0$, and $\Gamma_0=\vfi_0^{-1}(diag(\ome\cap U)).$
Let $(z,g)\in\Gamma_0$. If $\aaa\in T_z\ome$ and $A\in \alg$, then we have:
$$d_{(z,g)}\vfi_0(\aaa,Ag)=(\aaa,\mg \aaa+Az).$$
From (\ref{tgtR0}), we deduce that $Im(d_{(z,g)}\vfi_0)=T_{(z,gz)}R_0$. Thus $\vfi_0$ is a submersion,
which ends the proof of the lemma.\qed\\\\
-- {\it Calculus of the kernel of the Hessian of $\Phi$}:\\
Let $(z_0, g_0)\in\Gamma_0$. We define the following local chart of $\rdd\times G$ at $(z_0,g_0)$:
\beq\label{carte1}
\vfi(z,s):=(z,\exp(\sum_{i=1}^p s_i A_i) g_0).
\eeq
where $(A_1,\dots,A_p)$ is a  basis of $\alg$ that we should choose later.
We denote by:
\beq
\mbox{Hess } \Phi(z_0,g_0):=\left(\frac{\partial^2 (\Phi\circ \vfi)}{\partial x_i\partial x_j}(z_0,0)\right)_{1\leq i,j\leq 2d+p}
\eeq
the Hessian matrix in the canonical basis of $\rdd\times\rr^p$ of $\Phi\circ\vfi$. We clearly have:
$$\Phi^{\prime\prime}(z_0,g_0)_{|_{\mathcal{N}_{(z_0,g_0)}\Gamma_0}} \mbox{ is non degenerate } \iff
d_{(z_0,0)}\vfi (\ker_{_{\rr}} \mbox{Hess } \Phi(z_0,g_0)) \subset T_{(z_0,g_0)}\Gamma_0,$$
where $\ker_{_{\rr}} \mbox{Hess } \Phi(z_0,g_0):=\ker(\Re(\mbox{Hess } \Phi(z_0,g_0)))\cap \ker(\Im(\mbox{Hess } \Phi(z_0,g_0)))$.
To compute the matrix $\mbox{Hess } \Phi(z_0,g_0)$, we recall that:
\beq\label{liecalcul}
\frac{\partial^2}{\partial_{s_i}\partial_{s_j}}( e^{\sum_{r=1}^p s_r A_r})_{|_{s=0}} =\ud(A_i A_j + A_j A_i).
\eeq
After computation, if we denote $\mg$ by $g$, if `$i$' is associated to a line and `$j$' to a column, 
$i,j=1,\dots,p$, we get:
$$\mbox{Hess } \Phi(z,g)=\left(
\barr{c|c}
\ud J(g-\gmu) +\frac{i}{2}(I-g)(I-\gmu) & \ud [J(I+\gmu)+ i(\gmu -I) ]A_j z\\\hline
\ud^t[(J(I+\gmu)+ i(\gmu -I) )A_i z] & \frac{i}{2} <A_i z, A_j z>             \\
\earr\right).$$
To compute the  kernel of the Hessian of $\Phi$, we need to use the following formula:
\beq\label{form2}
\forall A\in\alg, \; \forall B\in\alg, \; \forall z\in\ome, \quad <Az, JBz>=0.
\eeq
Formula (\ref{form2}) comes from the fact that
$\ome$ is invariant by $G$ and is obtained by differentiating at $t=0$ the identity $F_A(e^{tB}z)=0$.
We set $x:=(g-I)\aaa+A z$. Then, we get:
\beq\label{ooo}
(\aaa,s)\in\ker_{_{\rr}}\mbox{Hess } \Phi(z,g)     \iff \left\{
\barr{l}
(I+\gmu) x=0\\
<(I+\gmu)JA_jz;\aaa>=0\quad \forall j=1,\dots, p\\
(\gmu-I)x=0\\
<A_j z;x>=0\quad \forall j=1,\dots, p
\earr\right.
\eeq
which is equivalent to $x=0$ and $(I+g)\aaa\in (J\alg z)^{\perp}$.
In addition, $x=0\iff (g-I)\aaa=-A z$, and, in view of (\ref{form2}), $Az\perp J\alg z$.
Thus (\ref{ooo}) $\iff x=0$ and $\aaa\in (J\alg z)^{\perp}$. Therefore:
\beq
d_{(z,0)}\vfi[\ker_{_{\rr}}\mbox{Hess } \Phi(z,g)]=\{ (\aaa ,Ag)\in\rdd\times\alg g : (M(g)-I)\aaa+Az=0
\mbox{ et } \aaa\in (J\alg z)^{\perp}\}. 
\eeq
According to Lemma \ref{varietefixe}, we have $d_{(z,0)}\vfi[\ker_{_{\rr}}\mbox{Hess }
\Phi(z,g)]=T_{(z,g)}\Gamma_0$. Hence, we have shown that there is  a theorical asymptotic expansion of
$\tr(f(\hqr))$. We have now to compute the first term and interpret it geometrically.
\subsection{Computation of the leading term}
Since $Q_0=1$ and $P_0=1$, we get from (\ref{viso}), (\ref{viso2}):
$$\tr(\widehat{A}_{\chi})\underset{h\to0^+}{\sim}(2\pi h)^{-d} d_{\chi}\int_G \gra
e^{\frac{i}{h}\Phi(z,g)}\carb a(z) dz dg.$$
By the generalized stationary phase theorem, we obtain from Lemma \ref{varietefixe}:
\beq\label{kukurucu}
\tr(\widehat{A}_{\chi})=(2\pi h)^{k_0-d} d_{\chi}\int_{\Gamma_0}
\frac{\carb a(z)}{\det_+^{\ud} \left(
\frac{\Phi^{\prime\prime}(z,g)}{i}_{|_{\mathcal{N}_{(z,g)}\Gamma_0}}
\right)} d\sigma_{\Gamma_0}(z,g) + O(h^{k_0-d+1}).
\eeq
where $d\sigma_{\Gamma_0}$ denotes the Riemannian measure on $\Gamma_0$. We have to compute the determinant
of this expression, and next we will give an integration formula to get from $\Gamma_0$ to $\omer$.\\\\
-- {\it Computation of the determinant of the transversal Hessian}:\\
Fix $z$ in $\rdd$ and $g$ in $G$ such that $\mg z=z$. We denote by $Stab(z)\subset G$ the stabilizer
of $z$. We endow the space of $d\times d$ matrices with the Riemannian structure coming from the following scalar
product:
\beq\label{scalarprod}
<<A,B>>:=\tr(^tAB), \quad \mbox{ for all matrices } A \mbox{ and }B.
\eeq
We choose a basis $A_1,\dots, A_p$ of $\alg$ such that:
\beq\label{orthocarte1}
A_1,\dots, A_{k_0} \mbox{ is an orthonormal basis of } (T_{Id}\, Stab(z))^{\perp} \mbox{ for } <<\, ;\,>>.
\eeq
\beq\label{orthocarte2}
A_{k_0+1},\dots, A_p \mbox{ is an orthonormal basis of }T_{Id}\, Stab(z) \mbox{ for } <<\,;\,>>. 
\eeq
By definition, we have:
$$\det\left(\Phi^{\prime\prime}(z,g)_{|_{\mathcal{N}_{(z,g)}\Gamma_0}}\right)=
\det((\Phi^{\prime\prime}(z,g)(\mu_i,\mu_j)))_{1\leq i,j \leq 2k_0}$$
where $(\mu_1,\dots,\mu_{2k_0})$ is an orthonormal basis of  $\mathcal{N}_{(z,g)}\Gamma_0$. We have:
$$\det\left(\Phi^{\prime\prime}(z,g)_{|_{\mathcal{N}_{(z,g)}\Gamma_0}}\right)=
\det\left(((<\mbox{Hess } \Phi(z,g)(d_{(z,0)} \vfi)^{-1}(\mu_i),(d_{(z,0)} \vfi)^{-1}(\mu_j)>))_{1\leq i,j
\leq 2k_0}\right).$$
Note that the differential of the chart $d_{(z,0)} \vfi$ is an isometry since $A_1,\dots, A_p$ is an
orthonormal basis of $\alg$. Thus, if $\e_i:=(d_{(z,0)} \vfi)^{-1}(\mu_i)$, then $(\e_1,\dots, \e_{2k_0})$
is an orthonormal basis of $(d_{(z,0)} \vfi)^{-1}(\mathcal{N}_{(z,g)}\Gamma_0)=\mathcal{F}^{\perp}$, where:
$$\mathcal{F}:= \{ (\aaa,s)\in\rdd\times \rr^p:(\mg-Id)\aaa+\sum_{i=1}^p s_iA_iz=0, \;\aaa\in T_z\ome
 \}.$$
By definition of $\Phi$, we note that  $\mbox{Hess } \Phi(z,g)(\mathcal{F}^{\perp})\subset
\mathcal{F}^{\perp}+ i\mathcal{F}^{\perp}$.
Hence, we have:
$$\det\left(\frac{\Phi^{\prime\prime}(z,g)_{|_{\mathcal{N}_{(z,g)}\Gamma_0}}}{i}\right)=
\det\left(\frac{\A_{|_{\mathcal{F}^{\perp}}}}{i}\right),$$
where $\A{|_{\mathcal{F}^{\perp}}}$ denotes the matrix of  the restriction of $\mbox{Hess } \Phi(z,g)$ to
$\mathcal{F}^{\perp}$ in {\it any} basis of $\mathcal{F}^{\perp}$.
We point out the fact that, differentiating the equality: $e^{tA}z=z$ at $t=0$,  we have, for $A\in\alg$:
\beq\label{form0}
Az=0 \iff A\in T_{Id}Stab(z).
\eeq
As a corollary, we note that $(A_1 z,\dots,A_{k_0} z)$ is a basis of $\alg z$.
The following lemma shows off a basis of $\mathcal{F}^{\perp}$:
\begin{lem}\label{base}
Let $(B_1z,\dots,B_{k_0} z)$ be a basis of $\alg z$.
We set in $\rdd\times\rr^p$, for $j=1,\dots,k_0$ :
$$\e_j:=(J B_j z,0),\qquad \e_j^{\prime}:=((\mgm-I) B_j z, <A_i z ,B_j z>,0) \quad (i=1,\dots,k_0).$$
Then
$\mathcal{B}:=(\e_1,\dots,\e_{k_0},\e^{\prime}_1,\dots,\e^{\prime}_{k_0})$
is a basis of $\mathcal{F}^{\perp}$.
\end{lem}
The proof is straightforward using (\ref{form2}), (\ref{orthocarte1}), (\ref{orthocarte2}), (\ref{form0})
and the fact that $T_z\ome=(J\alg z)^{\perp}$.
Calculating $\A_{|_{\mathcal{F}^{\perp}}}$ in this basis, a tedious but basic computation leads to:
$$\det\left(\frac{\A_{|_{\mathcal{F}^{\perp}}}}{i}\right)=\det\left(
\barr{c|c}
\ud(I-g_0)(I-g_0^{-1}) & \frac{1}{2i}(I+g_0^{-1})(g_0-I)(g_0^{-1}-I)+\frac{1}{2i}(I+g_0^{-1})M^tM\\\hline
\ud (I+g_0)            & \ud(g_0-I)(g_0^{-1}-I)+ \ud M^tM
\earr\right),$$
where $M$ is the $k_0\times k_0$ matrix with general term $<B_i z,A_j z>$ ($i,j=1,\dots,k_0$), and $g_0$ is the matrix of
the restriction of $\mg$ to $\alg z$ in the basis $(B_1 z,\dots,B_{k_0} z)$. Then, by the line operation $L_1\gets
L_1 -\frac{1}{i}(I+g_0^{-1}) L_2$, we get:
$$\det\left(\frac{\A_{|_{\mathcal{F}^{\perp}}}}{i}\right)=
\det[(I-g_0)(I-g_0^{-1}) +M^tM].$$
If one denotes by $f:\alg z\to \alg z$ the linear application defined by:
\beq\label{f}
f(x):=\sum_{r=1}^{k_0} <A_r z,x> A_r z, \quad \forall x\in \alg z,
\eeq
then we remark that the matrix of $f$ in the basis $(B_1 z,\dots,B_{k_0} z)$ is equal to $^tMM$. Therefore, we
have:
\beq
\det\left(\frac{\Phi^{\prime\prime}(z,g)}{i}_{|_{\mathcal{N}_{(z,g)}\Gamma_0}}\right)= 
\det[(I-M(g))(I-M(g)^{-1})_{|_{\alg z}} + f\,].
\eeq
-- {\it Integration formula on $\Gamma_0$}:
we have to deduce an integral over $\omer$ from an integral over $\Gamma_0$. A first step consists in passing
from $\Gamma_0$ to $\ome$. We will use the following integration lemma using submersions (a proof can
be found in \cite{these} or more generally in \cite{B-Z}):
\begin{lem}\label{lemmetranche}
Let $M$ and $N$ be two Riemannian manifolds, and $F:M\to N$ a smooth submersion. Let $\vfi$ be smooth
with compact support in $M$. Then we have:
$$\int_M \vfi(x)d\sigma_M(x) =\int_N \left[ \int_{\Sigma_n} \vfi(y)
 \frac{d\sigma_{\Sigma_n}(y)}{|\det(d_yF^t(d_y F))|^{\ud}}\right] d\sigma_N(n).$$
where $\Sigma_n:=F^{-1}(\{ n \})$, and $d\sigma_{\Sigma_n}$ is the Riemannian
measure induced by the one of $M$ on $\Sigma_n$.
\end{lem}
Let $\pi_1:\Gamma_0\to \ome\cap U$ be defined by: $\pi_1(z,g)=z.$
We recall that (see (\ref{tangentgamma})) if $\mg z=z$, then $\forall \aaa\in T_z\ome, \; (\mg-I)\aaa\in
\alg z$. From this fact, we deduce that $\pi_1$ is a submersion. Thus, we can apply Lemma
\ref{lemmetranche}, noting that $\Gamma_0$ is endowed with the Riemannian structure coming from the
scalar product (\ref{scalarprod}) and $G$ with the one of $d\tilde{g}$ Riemannian measure coming from
the same scalar product ($d\tilde{g}=Vol(G) dg$)\footnote{$Vol(G)$ denotes the volume of $G$ with respect to
the Riemannian structure of (\ref{scalarprod}). }. If we denote $Stab(z)$ by $H_z$, and if $\vfi$ is
smooth, compactly supported in $\Gamma_0$, then we have:
\beq\label{cocorico}
\int_{\Gamma_0} \vfi(z,g) d\sigma_{\Gamma_0}(z,g)=\frac{1}{Vol(G)} \int_{\ome}\left[ \int_{H_z} \vfi(z,y)
 \frac{d\tilde{\sigma}_{H_z}(y)}{|\det(d_{(z,y)}\pi_1. ^t(d_{(z,y)} \pi_1))|^{\ud}}\right] d\sigma_{\ome}(z),
\eeq
where $d\tilde{\sigma}_{H_z}$ also denotes the Riemannian measure from (\ref{scalarprod}) on $H_z$.
We recall that $H_0$ is the stabilizer of a fixed element $z_0$ of $\ome\cap U$.
If $u\in\mathcal{C}^{\infty}(H_z)$, and if $g_z\in G$ is such that $g_z H_0 g_z^{-1}=H_z$,
then it is easy to check that:
$$\int_{H_z} u(h) d\sigma_{H_z}(h) =\int_{H_0} u(g_z h g_z^{-1}) d\sigma_{H_0}(h),$$
where $d\sigma_{H_z}$ and $d\sigma_{H_0}$ are the normalized Haar measures on $H_z$ and $H_0$.
Therefore, using the fact that $Vol(H_z)=Vol(H_0)$ for (\ref{scalarprod}) (remember that $G$ is made of
isometries), we have, in view of (\ref{cocorico}):
\beq\label{sansev}
\int_{\Gamma_0} \vfi(z,g) d\sigma_{\Gamma_0}(z,g)= \int_{\ome}\mbox{{\scriptsize
$\frac{Vol(H_0)}{Vol(G)}$}}\left[
\int_{H_0} \vfi(z,g_z h g_z^{-1})
\frac{d\sigma_{H_0}(h)}{|\mbox{{\scriptsize $\det(d_{(z,g_z h g_z^{-1})}\pi_1. ^t(d_{(z,g_z h g_z^{-1})
} \pi_1))|^{\ud}$}}}
\right] d\sigma_{\ome}(z).
\eeq
\begin{lem}\label{croc} If $z\in\ome\cap U$ and $g\in G$ are such that $M(g)z=z$, then:
$$\det\left[d_{(z,g)}\pi_1. ^t(d_{(z,g)} \pi_1)\right]^{-1}=
\mbox{det}^{-1}(f)\det[(I-g)(I-g^{-1})_{|_{\alg z}} + f\,].$$
where $f$ is given by (\ref{f}).
\end{lem}
\underline{{\it Proof}} : first we remark that, if $g\in G$, then $T_z\ome=(J\alg z)^\perp$ is invariant by
$\mg$. Moreover, we note that, in view of (\ref{form2}), $\alg z\subset T_z\ome$. Let us denote by
$\mathcal{F}_0$ the orthogonal of $\alg z$ in $T_z\ome$. We note that $\mathcal{F}_0$ is also invariant
by $\mg$. Thus, if $\aaa\in \mathcal{F}_0$, then $(M(g)-I)\aaa\in \F_0$. But remembering that
$(M(g)-I)\aaa\in\alg z$, we get that $\mg=Id$ on $\F_0$.\

Besides, if $\aaa\in T_z\ome$, by definition of a transpose map, $^td_{(z,g)}\pi_1(\aaa)$ is the unique element
$(\aaa_0,A_0g)$ of $T_{(z,g)}\Gamma_0$ satisfying the fact that, for all $(\beta,Ag)\in T_{(z,g)}\Gamma_0$
we have:
\beq\label{mbaouafou}
<\beta, \aaa>_{\rdd}=<\beta,\aaa_0>_{\rdd} + <<A,A_0>>=0.
\eeq
Moreover, if $\aaa\in \F$, since $M(g)\aaa=\aaa$, we have $(\aaa,0)\in T_{(z,g)}\Gamma_0$.
Therefore, we have $^td_{(z,g)}\pi_1(\aaa)=(\aaa,0)$, and thus $d_{(z,g)}\pi_1.^td_{(z,g)}\pi_1=Id$ on
$\F_0$. Hence, we have:
$$\det(d_{(z,g)}\pi_1.^td_{(z,g)}\pi_1)=\det({d_{(z,g)}\pi_1.^td_{(z,g)}\pi_1}_{|_{\alg z}}).$$
Let us show that ${d_{(z,g)}\pi_1.^td_{(z,g)}\pi_1}_{|_{\alg z}}=(f+(I-M(g))(I-M(\gmu)))^{-1}
\circ f$. If $\aaa\in\alg z$, if we set $(\aaa_0,A_0 g):=^td_{(z,g)}\pi_1(\aaa)$, then we have to show
that:
$$f(\aaa)=(f+(I-M(\gmu))(I-M(g))) \aaa_0,$$
that is to say that (since  $(A_1 z ,\dots,A_{k_0} z )$ is a basis of $\alg z$), for
$i=1,\dots,k_0$, we have:
\beq\label{peterpan}
<A_i z,f(\aaa-\aaa_0)>=<A_i z,(I-\mgm)(I-\mg)\aaa_0>.
\eeq
The l.h.s. of (\ref{peterpan}) is equal to $\sum_{j=1}^{k_0} <A_i z, A_j z><\aaa-\aaa_0,A_j z>$.
Moreover, note that we have: $(A_j z, [A_j-gA_j\gmu] g)\in T_{(z,g)}\Gamma_0$. Thus, by
(\ref{mbaouafou}), we have:
\beq\label{eole}
 <A_jz, \aaa-\aaa_0>=<<(A_j-g A_j\gmu),A_0>>.
\eeq
Besides, 
$$<A_i z,(I-\gmu)(I-g)\aaa_0>=<A_i z,(I-\gmu)A_0 z>=<A_i z,(A_0-\gmu A_0 g) z>.$$
Then, decomposing $A_0-\gmu A_0 g$ in the orthonormal basis $(A_1 ,\dots,A_{k_0} )$ of $[T_{Id}Stab(z)]^{\perp}$,
we get:
$$<A_i z,(I-\gmu)(I-g)\aaa_0>=\sum_{j=1}^{k_0}<A_i z, A_j z> <<A_0-\gmu A_0 \, g,A_j>>.$$
From this last equality and (\ref{eole}), we get (\ref{peterpan}), which ends the proof of Lemma \ref{croc}.\qed\\\

We apply this lemma and (\ref{sansev}) to $\vfi(z,g):=\carb a(z)
\mbox{det}^{-\ud}\left(\frac{\Phi^{\prime\prime}(z,g)}{i}_{|_{\mathcal{N}_{(z,g)}\Gamma_0}}\right)$, to get:
\beq
\int_{\Gamma_0}\carb a(z) 
\mbox{{\scriptsize
 $\mbox{det}^{-\ud}\left(\frac{\Phi^{\prime\prime}(z,g)}{i}_{|_{\mathcal{N}_{(z,g)}\Gamma_0}}\right)$
 }}
 d\sigma_{\Gamma_0}(z,g)=\frac{Vol(H_0)}{Vol(G)}
\int_{\ome}\frac{a(z)}{\mbox{det}^{\ud}(f)} d\sigma_{\ome}(z).
\int_{H_0} \overline{\chi(h)} d\sigma_{H_0}(h).
\eeq
Note that, according to \cite{Se} or \cite{Si}, we have:
\beq
\int_{H_0} \overline{\chi(h)} d\sigma_{H_0}(h)=\left[{\rho_{\chi}}_{|_{H_0}}:\indic\right].
\eeq
Finally, we are greatly indebted to Gilles Carron for giving us the proof of the following lemma:
\begin{lem}\label{gillescarron}
$\mbox{det}^{\ud}(f)=Vol(G(z)) \frac{Vol (H_0)}{Vol(G)}$, where $Vol (G)$ and $Vol (H_0)$ are calculated
as Riemannian volumes for the scalar product (\ref{scalarprod}).
\end{lem}
\underline{{\it Proof}} : let  $(\e_1,\dots,\e_{k_0})$ be an orthonormal basis of $\alg z$. We have:
\beq
\det(f)=\det[((<A_i z, \e_j>))_{(1\leq i,j\leq k_0)}]^2.
\eeq
The idea of the proof is to use the link between $G$ and $G(z)$ through $G/H$, using the fact that there is a
unique $G$-invariant volume form\footnote{A volume form on $G/H$ is said to be $G$-invariant if we have for
all $g$ in $G$: $L_g^*\aaa=\aaa$, where $L_g:G/H\to G/H$ is defined by $L_g(g_0H)=(gg_0)H$ for $g_0\in G$.} on
$G/H$ (up to a constant).
We denote here by $e$ the neutral element of $G$ ($e=Id$), $H:=Stab(z)$ and $\h:=T_eH$. Let $\pi:G\to G/H$
be the canonical projection on the quotient. Via $\pi$, $G/H$ inherits a Riemannian structure from $G$.
Namely, if $g$ is any element of $G$, let $\psi_g$ denote the reciprocal application of the restriction of
$d_g\pi$ to $(\ker d_g\pi)^{\perp}=(g\h)^\perp$. Then the metric $\delta$ on $G/H$ is defined by:
$$\forall g\in G,\; \forall u,v \in T_{\pi(g)}(G/H),\quad\delta_{\pi(g)}(u,v)=<<\psi_g(u),\psi_g(v)>>.$$
Let $\omega$ be the associated Riemannian volume form on $G/H$. We claim that:
\beq\label{volform1}
\left\{
\barr{l}
\omega \mbox{ is $G$-invariant.}\\
\omega(eH)(\psi_e^{-1}(A_1),\dots ,\psi_e^{-1}(A_{k_0}))=1.\\
\int_{G/H}\omega=\frac{Vol G}{Vol H}.
\earr\right.
\eeq
(use Lemma \ref{lemmetranche} and the fact that $\Psi_e$ is an isometry).\

Another volum form on $G/H$ is given by $\Phi^*\mu$, where $\Phi:G/H\to G(z)$ is defined by $\Phi(gH)=gz$, for
any $g$ in $G$, and $\mu$ denotes the volume form associated to the euclidian volume on $G(z)$. We claim that:
\beq\label{volform2}
\left\{
\barr{l}
\Phi^*\mu \mbox{ is $G$-invariant.}\\
\Phi^*\mu(eH)(\psi_e^{-1}(A_1),\dots ,\psi_e^{-1}(A_{k_0}))=1.\\
\int_{G/H}\Phi^*\mu=Vol \, G(z).
\earr\right.
\eeq
Indeed, for $g\in G$, $\Phi\circ L_g=g.\Phi$, and $\mu$ is $G$-invariant. Moreover, 
$\psi_e^{-1}(A_i)=d_e\pi(A_i)$, and since $\Phi\circ \pi(g)=gz$, we have $d_{eH}\Phi \circ d_e\pi(A_i)=A_i z$
for $i=1,\dots,k_0$.\

Now, if $\Omega^{k_0}(G/H)^G$ denotes the set of all $G$-invariant volume forms of $G/H$, we note that the
application $\aaa\mapsto \aaa(eH)$ from $\Omega^{k_0}(G/H)^G$ to $\Lambda^{k_0}(T_{eH}(G/H))$
is an isomorphism. In particular $\dim(\Omega^{k_0}(G/H)^G)=1$, and thus we deduce that there exists
$\la$ in $\rr$ such that:
$$\Phi^*\mu=\la\, \omega.$$
By (\ref{volform1}) and (\ref{volform2}), we obtain that $\la=Vol(G(z)) \frac{Vol(H)}{Vol(G)}$, and, besides,
 that
$\la=\det(\e_i,A_j z)=(\det(f))^{\ud}$.\qed\\\

Thus, in view of (\ref{kukurucu}):
$$\tr(\widehat{A}_{\chi})=(2\pi h)^{k_0-d} d_{\chi}\int_{\ome}a(z)
\frac{d\sigma_{\ome}(z)}{Vol(G(z))} \left[{\rho_{\chi}}_{|_{H_0}}:\indic\right] \; +\; O(h^{k_0-d+1}).$$
where $dh$ is the Haar measure on $H_0$. The proof is clear if we remark that:
if $v\in \mathcal{C}_0^{\infty}((\ome\cap U)/G)$, we have, using Lemma  \ref{lemmetranche}:
\beq\label{intomer}
\int_{\omer} v(x) d\sigma_{red}(x) = \int_{\ome} v(\pi(z)) \frac{d\sigma_{\ome}(z)}{Vol(G(z))}.
\eeq
\section{Reduced Gutzwiller formula: $G$-clean flow conditions}
We now focus on the asymptotics of the reduced spectral density (\ref{trass}) (see Introduction for
notations). The case where $\hat{f}$ is supported near zero will lead to Theorem \ref{Comptage}, and the case
where $0\notin \supp\hat{f}$ to Theorem \ref{Gutz}. By formula (2.3) of \cite{fini} and linearity of the
trace, we have:
$$\mathcal{G}_{\chi}(h):=d_\chi\int_G \carb \; Tr\left(\psi(\hq) f\left( \frac{E-\hq}{h} \right) \mm \right)\,dg.$$
Note that the proof of \cite{fini}, in particular section 3 (`Reduction of the proof by coherent states')
gives us an asymptotic expansion at fixed $g$ of the quantity $I_g(h):=Tr\left(\psi(\hq) f\left( \frac{E-\hq}{h}
\right) \mm \right)$ for which the rest is uniform with respect to  $g\in G$, since $G$ is compact. Namely, we made
hypothesis (\ref{hypCF}) on $H$ in order to have a pleasant description of term $\psi(\hq)$ by functional
calculus.
Under this assumption, using the theorem of propagation of coherent states given by Combescure and Robert in
\cite{CR} and \cite{Ro1}, we found that $I_g(h)$ has an asymptotic expansion in powers of $h$ with
coefficients depending on $h$ and of the form:
$$\int_{\rr_t}\int_{\rdd_z} \exp\left(\frac{i}{h}\vfi_E(t,z,g)\right) \hat{f}(t) a_g(z) dz dt.$$
where $a_g:\rdd\to\cc$ is supported in $H^{-1}(]E-\delta E,E+\delta E[)$, and $\vfi_E=\vfi_1+i \vfi_2$ with:
\beq\label{phase1}
\left\{
\barr{l}
\disp{ \vfi_1(t,z,g):=(E-H(z))t+\ud<M(g)^{-1}z,Jz>-\ud\int_0^t(z_s-\mgm z)J\dot{z_s}ds}\\
\disp{ \vfi_2(t,z,g):=\frac{i}{4}<(I-\wt)(\mg z_t-z);(\mg z_t-z)>.}
\earr\right.
\eeq
where $z_t:=\Phi_t(z)$, $\Phi_t$ being the flow of (\ref{hamsyst}).
We set 
$F_z(t):=\partial_z \Phi_t(z)\in Sp\,(d,\rr)$ and 
\beq\label{monodr}
F_{z}(t)=\left(
\begin{array}{cc}
A & B \\
C & D \\
\end{array}
\right)
\qquad\mbox{ where }\; A, B, C, D \mbox{ are $d\times d$ matrices}.
\eeq
Then $\widehat{W_t}:=\left(
\barr{cc}
W_t      & -iW_t\\
-i\, W_t & -W_t
\earr
\right)$ with $\ud (I+W_t):=(I-i^tg^{-1}\, M_0\gmu)^{-1}$, where we have set $M_0:=(C+iD)(A+iB)^{-1}$.
Moreover, we have from \cite{fini} that $\norm{W_t}_{\li(\cc^{d})}<1$.
Therefore, $\mathcal{G}_{\chi}(h)$ has an asymptotic expansion in powers of $h$ with
coefficients depending on $h$ and of the form:
\beq\label{J(h)}
J(h)=\int_G\int_{\rr_t}\int_{\rdd_z} \exp\left(\frac{i}{h}\vfi_E(t,z,g)\right) \hat{f}(t) a_g(z)
dz \, dt\, dg.
\eeq
where $a_g:\rdd\to\cc$ is supported in $H^{-1}(]E-\delta E,E+\delta E[)$. In particular, in view of
\cite{fini}, we get:
\beq\label{firstterm}
\mathcal{G}_{\chi}(h)\underset{h\to 0^+}{\sim} d_{\chi}\frac{(2\pi h)^{-d}}{2\pi}
\int_G \int_{\rr_t}\int_{\rdd_z}\carb  \chi_1(z) \hat{f}(t)
\psi(H(z))\mbox{ {\scriptsize $ \detp( \frac{A+iB-i(C+iD)}{2})$} }  
e^{\frac{i}{h}\varphi_{E}(t,z,g)}dtdz dg,
\eeq
where $\chi_1$ is a smooth function compactly supported in $\rdd$, equal to $1$ on $\nrj=\{H=E\}$. Note that
problems of entire powers of $h$ and of shifting in powers of $h$ are solved with exactly the same
method as in \cite{fini}.
We have now to apply the generalised stationary phase theorem (in the version of \cite{CRR}, Theorem 3.3), to
each term $J(h)$. In this section, we describe the minimal hypotheses to be fulfilled for applying this
theorem. These will be called $G$-clean flow assumptions. We then give the theorical asymptotics in Theorem
\ref{asymptheolie}. We will compute leading terms of the expansion in two particular cases in sections 5
and 6. For the proof of Theorems \ref{asymptheolie}, \ref{Weyl}, \ref{Gutz}, we recall that, by an averaging argument, we can
suppose that the group $G$ is made of isometries (see \cite{fini}). We will often denote $\mg$ by $g$, in
order to simplify notations. Finally, we suppose that hypotheses of symplectic reduction are fulfilled in
$\ome\cap U$, where $U:=H^{-1}(]E-\delta E,E+\delta E[)$.
\subsection{Computation of the critical set}
Let $$\ce:=\{ a\in \rr\times \rr^{2d}\times G : \Im (\varphi_E(a))=0,
\nabla\varphi_E(a)=0 \}.$$
\begin{prop}\label{calculenscrit}
The critical set is:
\begin{equation}\label{critik}
\ce=\{(t,z,g)\in \rr\times \rdd\times G : z\in(\ome\cap \Sigma_E),
\mg \Phi_t(z)=z \}.
\end{equation}
\end{prop}
\underline{{\it Proof:}} as in \cite{fini}, using that $\norm{W_t}_{\li(\cc^d)}<1$, we get that
$\Im \vfi_E(t,z,g)=0 \iff \mg\Phi_t(z)= z$.\\
We first need some formulae coming from the symmetry that will be helpful for the computation:
We recall that $F_z(t)=\partial_z(\Phi_t(z))$. By differentiating formula (\ref{Ginvariant}), we get:
\beq\label{grad}
\nabla  H(\mg z)=^t\mgm \nabla H(z),\qquad \forall z\in\rdd,\;\forall g\in G.
\eeq
This formula implies that we have also:
\beq\label{flowsym}
\Phi_t(\mg z)=\mg \Phi_t(z), \; \forall z\in\rdd,\;\forall g\in G,\; \forall t\in\rr \mbox{ such that the flow exists at time $t$}.
\eeq
Moreover we recall that, since $\mg$ is symplectic, we have:
\beq
J\mg=^t\mgm J \;\mbox{ and }\; \mg J =J^t\mgm.
\eeq
Finally, if $t$ and $z$ are such that $\mg \Phi_t(z)=z$, then we have:
\beq\label{vecpropre}
(\mg F_z(t)-I)\jgrad=0 \mbox{ and } (^tF_z(t)^t\mg -I)\grad=0.
\eeq
The second identity comes from the first one since $\mg F_z(t)$ is symplectic. For this first identity, one
can differentiate at $s=t$ the equation:
$$\Phi_t(\mg\Phi_s(z))=\Phi_s(z).$$
Moreover, we recall (\ref{form2}), and claim that we have: 
\beq\label{form1}
\forall A\in\alg, \; \forall z\in\ome, \quad <Az, \grad>=0.
\eeq
Formula (\ref{form1}) is equivalent to say that $\{H,F_A\}=0$.\\
$\bullet$ {\it Computation  of the gradient of $\varphi_1$}: if $A\in\alg$, then we have:
\beq\label{deriv1}
\left\{
\barr{l}
\partial_t\vfi_1(t,z,g)=E-H(z)-\ud<(z_t-\mgm z);J\dot{z}_t>\\
\nabla_z\vfi_1(t,z,g)=\ud (\mg+^tF_z(t))J(z_t-\mgm z)\\
\partial_g \vfi_1(t,z,g)(Ag)=\ud <JAz;\mg z_t>
\earr
\right.
\eeq
$\bullet$ {\it Computation  of the gradient of $\varphi_2$}:
\beq\label{deriv2}
\left\{
\barr{l}
4\partial_t\vfi_2(t,z,g)=2<(I-\widehat{W_t})(z_t-\gmu z);\dot{z}_t>-
<\partial_z(\widehat{W_t})(z_t-\gmu z);(z_t-\gmu z)>\\
4\nabla_z\vfi_2(t,z,g)=2(^tF_z(t)-g)(I-\widehat{W_t})(z_t-\gmu z)-
^t[\partial_z(\widehat{W_t})(z_t-\gmu z)](z_t-\gmu z)\\
4\partial_g \vfi_2(t,z,g)(Ag)=2<(I-\widehat{W_t})\gmu A z;(z_t-\gmu z)>,\quad \mbox{ if $A\in\alg$.}
\earr
\right.
\eeq
The only difficulty lies in the computation of $\nabla_z\vfi_1(t,z,g)$: we have: $\nabla_z\vfi_1(t,z,g)=$
$$-t\grad+\ud(gJ+^t(gJ))-\ud
\int_0^t \; ^t\partial_z(\Phi_s(z)-\gmu z) J\dot{z_s}ds-\ud \int_0^t \; ^t(J \partial_z(\dot{z_s}))
(z_s-\gmu z) ds.$$
We note that: $\partial_z(\dot{z_s})=\partial_z\partial_s(\Phi(s,z))=\frac{d}{ds}(F_z(s))$.
By an integration by parts, we obtain:
$$\int_0^t \;^t(J \partial_z(\dot{z_s})) (z_s-\gmu z)ds
=-\left[ ^t F_z(s) J(z_s-\gmu z)  \right]_0^t + \int_0^t ~^tF_z(s) J \dot{z_s} ds.$$
The end of the calculus is straightforward, if we note that:
$$^tF_z(s) J \dot{z_s}=- ^tF_z(s) \nabla H(\Phi_s(z))=-\nabla(
H(\Phi_s(z)) )=-\grad.$$
Therefore, we remark that $(t,z,g)\in \ce$ if  and only if $\Phi_t(z)=\mgm z$, $H(z)=E$ and for all $A$ in
$\alg$, $<JAz;z>=0$, i.e. $z\in \ome$. This ends the proof of proposition \ref{calculenscrit}.\qed
\subsection{Computation of the Hessian $\mathbf{\mbox{Hess} \;\vfi_{E}(t,z,g)}$}
The space of complex matrices is endowed with the scalar product (\ref{scalarprod}). If
$(t_0,z_0,g_0)\in\ce$, then we choose the chart of $\rr\times\rdd\times G$ at $(t_0,z_0,g_0)$ to be:
\beq\label{carte}
\varphi_{(t_0,z_0,g_0)}=\varphi : \left\{
\barr{c} \mathcal{U}\subset \rr\times\rdd\times \rr^p \to \rr\times \rdd\times G\\
\disp{(t,z,s)\mapsto (t,z,\exp(\sum_{i=1}^p s_i A_i). g_0).}
\earr\right.
\eeq
where $(A_1,\dots,A_p)$ is the orthonormal basis of  $\alg$  defined such that:
\beq\label{bonai1}
(A_1 ,\dots, A_{k_0}) \mbox{ is the orthonormal basis of   } [T_{Id}(Stab(z_0))]^{\perp}.
\eeq
\beq\label{bonai2}
(A_{k_0+1},\dots,A_p) \mbox{ is the orthonormal basis of } T_{Id}(Stab(z_0)).
\eeq
Then we define:
$$\mbox{Hess } \vfi_E(t_0,z_0,g_0):=\left(\left(\frac{\partial^2 (\vfi_E\circ \varphi)}{\partial x_i \partial
x_j}(t_0,z_0,0)\right) \right)_{1\leq i,j\leq p}  \in M_{2d+p+1}(\cc).$$
\begin{prop}\label{calculhess}
Let $(t,z,g)\in\ce$. Then $\mbox{Hess}\; \vfi_E(t,z,g)=$
$${\scriptsize \left(
\barr{c|c|c|c}
\frac{i}{2}<g(I-\wt)\gmu J\nabla H;J\nabla H> & -^t\nabla H &
\frac{i}{2}<J\nabla H;g(I-\wt)\gmu A_j z>  & 0\\
 & + \frac{i}{2}^t\left[(^tF-g)(I-\wt)\gmu J\nabla H\right] & & \\\hline

-\nabla H & \ud[Jg F-^t(g F)J]
 &  \ud (^tF+g)J\gmu A_j z & 0 \\
+\frac{i}{2}(^tF-g)(I-\wt)\gmu J\nabla H & +\frac{i}{2}(^tF-g)(I-\wt)(F-\gmu) &
+ \frac{i}{2}(^tF-g)(I-\wt)\gmu A_j z &  \\\hline

\frac{i}{2}<J\nabla H;g (I-\wt)\gmu A_i z> & \ud ^t\left[(^tF+g)J\gmu A_i z \right] &
\frac{i}{2}<g (I-\wt)\gmu A_i z;A_j z>& 0\\
 & +\frac{i}{2}^t[(^tF-g)(I-\wt)\gmu A_i z] & & \\\hline
 
0 & 0 & 0 & 0 
\earr
\right)}$$
where, we denoted $F_z(t)$ by $F$, and $\grad$ by $\nabla H$, and each line index $i$ and column index $j$
is repeated $k_0$ times.
\end{prop}
\underline{{\it Proof:}} use (\ref{deriv1}) and (\ref{deriv2}) together with (\ref{grad}), (\ref{flowsym}) and
(\ref{vecpropre}). For $\partial_{s_i}\partial_{s_j}\vfi_E$, use (\ref{liecalcul}) to write:
$$\frac{\partial^2}{\partial_{s_i}\partial_{s_j}}\varphi_1(t,z,g)=\frac{1}{4} <z;J(A_i A_j+ A_j A_i)gz> +
\frac{1}{4} \int_0^t <\gmu (A_i A_j+ A_j A_i)z, J\dot{z_s}> ds$$
$$=\frac{1}{4}<(A_i A_j+ A_j A_i)z,Jz>= -\ud<A_j z, JA_i z>=0,$$
in view of (\ref{form2}).
Besides, we have: $4\, \partial_{s_i}\partial_{s_j}\varphi_2(t,z,g)=$
$$\mbox{ {\scriptsize $\frac{d}{ds_j}_{|_{s_j=0}} \left[ <(I-\wt) \frac{d}{ds_i}_{|_{s_i=0}} (z_t-\gmu z) ;z_t-\gmu z>
+<(I-\wt)(z_t-\gmu z) ;\frac{d}{ds_i}_{|_{s_i=0}} (z_t-\gmu z)>\right]$}}$$
$$\mbox{ {\scriptsize $= < (I-\wt)\frac{d}{ds_i}_{|_{s_i=0}} (z_t-\gmu z) ; \frac{d}{ds_j}_{|_{s_j=0}} (z_t-\gmu z) >
+<(I-\wt)\frac{d}{ds_j}_{|_{s_j=0}} (z_t-\gmu z) ;
\frac{d}{ds_i}_{|_{s_i=0}} (z_t-\gmu z) >$ }} $$
$$=2<(I-\wt)\gmu A_i z,\gmu A_j z>.$$
If one remembers (\ref{form0}), then the proof of the proposition is clear.\qed
\subsection{Computation of the real kernel of the Hessian}
We denote here by $\ker_{_{\rr}}\mbox{Hess}\; \vfi_E(t,z,g)$ the set of $(\tau,\aaa,A g)$ in
$\rr\times \rr^{2d}\times \alg g$ such that $\disp{A=\sum_{i=1}^p s_i A_i}$ ($s_i\in\rr$)
and $(t,z,s_1,\dots s_p)\in \ker (\mbox{Hess} \,\vfi_E(t,z,g))$. We recall that, if $(t,z,g)\in\ce$, then:
$$\vfi_E''(t,z,g)_{|_{\mathcal{N}_{(t,z,g)}\ce}} \mbox{ is non degenerate } \iff \ker_{_{\rr}} (\mbox{Hess}\,
\vfi_E(t,z,g))\subset T_{(t,z,g)}\ce.$$
\begin{prop}\label{calculnoyau}
Let $(t,z,g)\in \ce$. Then $\ker_{_{\rr}}\mbox{Hess}\; \vfi_E(t,z,g)=$
\begin{equation}\label{noyau}
\{(\tau,\aaa,A g)\in \rr\times \rr^{2d}\times \alg g : \aaa \perp \nabla 
H (z), \aaa\perp J(\alg z), \tau J\nabla H(z)+(\mg F_z(t)-Id)\aaa +A z=0 \}.
\end{equation}
\end{prop}
\underline{{\it Proof:}} we set $\wt=:\wtu+i\wtd$, where $\wtu$ and $\wtd$ are the real and (resp.) imaginary
parts of $\wt$ ($\wtu$ and $\wtd$ are symmetric matrices). We denote $F_z(t)$ by $F$. Then, by
proposition \ref{calculhess}, if we set $x:=\tau \jgrad + (gF-I)\aaa+A z$, then $(\tau,\aaa,Ag)
\in\ker_{\rr}\mbox{Hess}\,\vfi_E(t,z,g)$ if and only if:
\beq\label{kernoyau}
\left\{
\barr{l}
<g\wtd \gmu \jgrad;x>=2<\grad;\aaa>.\\
<g(I-\wtu)\gmu \jgrad;x>=0.\\
(^tF-g)(I-\wtu)\gmu x =0. \qquad \qquad\qquad\qquad\qquad\qquad(\star)\\
<g\wtd \gmu A_i z;x> + <J A_i z;(gF+I)\aaa>=0, \; \mbox{ for $i=1,\dots, k_0$}.\\
<g(I-\wtu)\gmu A_i z;x>=0 \; \mbox{ for $i=1,\dots, k_0$}.
\earr\right.
\eeq
and
\beq\label{pique}
\left.
\barr{r}
-2\tau \grad  + [JgF-^t(gF)J]\aaa + (^tF-g)\wtd[(F-\gmu)\aaa+Az]\\
+\tau (^tF-g)\wtd\gmu \jgrad + (^tF+g)J\gmu Az=0.
\earr
\right\}
\eeq
We note that $\wtd=J\wtu$, $[g,J]=0$ and that $gF$ is a symplectic matrix. By multiplying the last equality
by $gFJ$ and using the fact that: $gF \jgrad=\jgrad$, we get:
$$(\ref{pique}) \iff (gF-I)(I-g\wtu \gmu)x=-2x.$$
We set $y:=(I-g\wtu \gmu) x.$ Then:
$$\left\{\barr{l}
(\star)\iff y\in \ker[^t(gF)-I]=Im(gF-I)^{\perp}.\\
(\ref{pique})\iff (gF-I) y=-2x.
\earr\right.$$
If $(\tau,\aaa,g A)\in \ker_{\rr} \mbox{Hess} \; \vfi_E (t,z,g)$ then $x\perp y$, i.e.
$$<(I-g\wtu\gmu)x;x>=0,\; \mbox{ i.e. }\; |x|^2=<\wtu\gmu x;\gmu x>.$$
Since $\norm{\wtu}<1$, we get $x=0$. Then, using  (\ref{kernoyau}), we get $\aaa\perp \grad$ and
$(gF+I)\aaa\in J(\alg z)^{\perp}$.
In view of (\ref{form1}) and (\ref{form2}), $\jgrad$ and $A z$ are in $(J\alg z)^{\perp}$, and $x=0$, thus:
$(gF-I)\aaa\in (J\alg z)^{\perp}$. Therefore, we have $\aaa\in (J\alg z)^{\perp}$.
The converse is clear.\qed
\subsection{Asymptotics under $G$-clean flow conditions}
We now give a simple geometrical criterium to have an asymptotic expansion of $\mathcal{G}_{\chi}(h)$ if
$T>0$ is such that $\supp(\hat{f})\subset ]-T,T[$:\

Let $\Psi:=\left\{
\barr{l}
]-T,T[\times (\nrj\cap\ome)\times G\to \rdd\\
(t,z,g)\mapsto \mg \Phi_t(z)-z
\earr\right.$
\begin{defi}\label{flop}
We say that {\it the flow is $G$-clean} on $]-T,T[\times (\nrj\cap\ome)$ if zero is a weakly regular value
of $\Psi$, i.e. :
\begin{itemize}
\item $\Psi^{-1}(\{ 0 \})=:\mathcal{C}_{E,T}$ is a finite union of submanifolds of $\rr\times\rdd\times G$.
\item $\forall (t,z,g)\in \mathcal{C}_{E,T}$, we have $T_{(t,z,g)}\mathcal{C}_{E,T} =\ker d_{(t,z,g)}\Psi$.
\end{itemize}
\end{defi}
\noindent If there is no critical point of $\widetilde{H}$ on $\nrjt$ (see (\ref{objetsreduits})), then the
$G$-clean flow hypothesis is somehow optimal to apply the generalised stationary phase theorem. Indeed,
 if $(t,z,g)\in\mathcal{C}_{E,T}$, then:
\beq\label{noyaupsiphi}
\ker d_{(t,z,g)}\Psi=\ker_{_{\rr}}\mbox{Hess} \,\vfi_E(t,z,g).
\eeq
The justification is the following: we first note that, if $(\tau,\beta,A)\in\rr \times
T_z(\ome\cap\nrj)\times \alg$, then:
$$d_{(t,z,g)}\Psi(\tau,\beta,A g)=\tau J\nabla H(z)+(\mg F_z(t)-Id_{\rdd})\beta+Az.$$
Moreover, we have the following lemma:
\begin{lem}\label{transversecrit}
We recall that hypotheses of symplectic reduction are fulfilled on $U\cap \ome$ where $U:=H^{-1}(]E-\delta E,
E+\delta E[)$. Then following assertions are equivalent:
\ben
\item $\ome\cap U$ and $\nrj$ are transverse submanifolds of $\rdd$.
\item There is no critical point of $\widetilde{H}$ on $\nrjt$.
\een
\end{lem}
Therefore, by Proposition \ref{calculnoyau}, we have (\ref{noyaupsiphi}).\\
\underline{{\it Proof of the lemma:}} the negation of $(1)$ is: $\exists z\in(\ome\cap U)\cap\nrj$, 
$T_z\ome\subset (\rr\nabla H(z))^{\perp}$, i.e., since $(J\mathcal{G}z)^{\perp}=T_z\ome$:
$\rr\nabla H(z)\subset J\mathcal{G}z$ that is $J\nabla H(z)\in\mathcal{G}z$. Finally, if $\pi:\ome\to\omer$
denotes the canonical projection on the quotient, then we have (since $\ker d_z\pi=\alg z$):
$d_{\pi(z)}\widetilde{H}=0$ if and only if $\jgrad \in \alg z$.\qed\\\\
We get the following theorem:
\begin{theo}\label{asymptheolie} 
Let $G$ be a compact Lie group of $Gl(d,\rr)$ and  $H:\rdd\to \rr$ be a smooth $G$-invariant Hamiltonian 
satisfying (\ref{hypCF}). Let $E\in\rr$ be such that $H^{-1}([E-\delta E, E+\delta E])$ is compact for some $\delta E>0$, and that
$\nrjt=\{ \widetilde{H}=E\}$ has no critical points.
We suppose that hypotheses of reduction are satisfied on $\ome\cap U$, where $U:=H^{-1}(]E-\delta E, E+\delta E[)$.
Let $f$ and $\psi$ be real functions in $\mathcal{S}(\rr)$ with $\supp(\psi)\subset ]E-\delta E,E+\delta E[$
and $\hat{f}$ compactly supported in $]-T,T[$ where $T>0$. Moreover, we suppose that the $G$-clean flow
conditions are satisfied on $]-T,T[\times(\ome\cap\nrj)$.
We denote by:
$$\mathcal{C}_{E,T}:=\{(t,z,g)\in ]-T,T[\times \rdd\times G : z\in(\ome\cap \Sigma_E), \mg \Phi_t(z)=z \},$$
and by  $[\mathcal{C}_{E,T}]$ the set of its connected components. Then the quantity $\int_0^t p_s \dot{q_s}ds$
($(q_t,p_t):=\Phi_t(z)$) is constant
on each element $Y$ of $[\mathcal{C}_{E,T}]$, denoted by $S_{Y}$ and $\mathcal{G}_{\chi}(h)$ has the following asymptotic
expansion modulo $O(h^{+\infty})$:
$$\mathcal{G}_{\chi}(h)=\sum_{Y\in [\ce]} (2\pi h)^{\frac{1-\dim Y +p}{2}} e^{\frac{i}{h}S_{Y}}\psi(E)
\frac{d_{\chi}}{2\pi}
\left( \int_Y \hat{f}(t) \carb d(t,z,g) d\sigma_Y(t,z,g) +\sum_{j\geq 1}h^j a_{j,Y}\right).$$
where the $a_{j,Y}$ are distributions in $\hat{f}$, and the density $d(t,z,g)$ is defined by:
$$d(t,z,g):= \detp\left( \frac{\varphi_{E}^{\prime\prime}(t,z,g)_{|_{\mathcal{N}_{(t,z,g)}Y}}}{i} \right)
\detp\left( \frac{A+iB-i(C+iD)}{2}\right).$$
\end{theo}
We recall that $\varphi_{E}^{\prime\prime}(t,z,g)$ is given by proposition \ref{calculhess} and
that $A$, $B$, $C$, $D$ are given by:
$\partial_z(\Phi_t(z))=F_z(t)=${\scriptsize $\left(
\barr{cc}
 A & B \\
 C & D
\earr\right).$}\\\\
\underline{{\it Proof of the theorem:}} we can apply the stationary phase theorem to each coefficient
(\ref{J(h)}). Then one can use (\ref{firstterm}) to calculate the first term, and remark that, on $\mathcal{C}_{E,T}$,
$\vfi_E$ is constant with value $\varphi_{E}(t,z,g)=\int_0^tp_s \dot{q_s}ds$, if $(t,z,g)\in\ce$.\qed
\section{The Weyl part}\label{Weylpart}
In this section, we plan to give asymptotics of $\mathcal{G}_{\chi}(h)$ when $\supp(\hat{f})\cap
\mathcal{L}_{red}(E)=\{ 0 \}$, where:
\beq
\mathcal{L}_{red}(E)=\mathcal{L}_{red}:=\{ t\in \rr:\exists g\in G,\exists z\in\ome\cap \nrj :  \mg\Phi_t(z)=z\}.
\eeq
Note that $\mathcal{L}_{red}(E)$ is the set of periods of periodic orbits of $\nrjt$. In particular, this
assumption is fulfilled if $\hat{f}$ is supported close enough to zero, which will lead to
Theorem \ref{Comptage}. 
\begin{prop}\label{cleanzero}
Suppose that $\nrjt$ is non critical. Then $0$ is isolated in $\mathcal{L}_{red}(E)$. Moreover, if
$\supp(\hat{f})\cap \mathcal{L}_{red}(E)=\{ 0 \}$, then $\ce\cap (\supp(\hat{f})\times\rdd\times G)=\{ 0
\}\times W_0$, where
\beq
W_0:=\{ (z,g)\in (\ome\cap\nrj)\times G : \mg z=z\}.
\eeq
$W_0$ is a submanifold of $\rdd\times G$, $\dim W_0=2d-2k_0+p-1$ and if $(z,g)\in W_0$, then
\beq
T_{(z,g)}W_0=\{ (\aaa, Ag)\in\rdd\times \alg g : (\mg-I)\aaa+Az=0, \aaa\perp \nabla H(z) \mbox{ and } \aaa \in T_z\ome \}.
\eeq
\end{prop}
\underline{{\it Proof:}} $\nrjt$ is a compact and non critical energy level. So it is a well known fact that
it has a minimal strictly positive period. If $\supp(\hat{f})\cap \mathcal{L}_{red}(E)=\{ 0 \}$, by the non
stationary phase theorem, our critical set becomes $\ce\cap (\supp(\hat{f})\times\rdd\times G)=\{ 0
\}\times W_0$. We note that $W_0=\Gamma_0\cap[(\nrj\cap\ome)\times G]$, where $\Gamma_0$ is given by
(\ref{Gamma_0}). By Lemma \ref{varietefixe} (but here with $U=H^{-1}(]E-\delta E,E+\delta E[)$), we know
that $\Gamma_0$ is a submanifold of $\rdd\times G$.
We have to show that $\Gamma_0$ and $(\nrj\cap\ome)\times G$ are transverse submanifolds of
$(\ome\cap U)\times G$: in the case of the contrary, let $(z,g)\in W_0$ such that
$T_z\Gamma_0\subset (T_z\nrj\cap T_z\ome)\times \alg g$.
If $\aaa\in T_z\ome$, we have seen in section 3 (\ref{tgtR0}) that $(\mg-I)\aaa\in\alg z$.
Hence, there exists $A\in\alg$ such that $(\aaa,Ag)\in T_{(z,g)}\Gamma_0$. Thus, we have
$T_z\ome\subset T_z\nrj$, which is in contradiction with lemma \ref{transversecrit}.\qed\\
\begin{theo}\label{Weyl}
Let $G$ be a compact Lie group of $Gl(d,\rr)$ and  $H:\rdd\to \rr$ be a smooth $G$-invariant Hamiltonian 
satisfying (\ref{hypCF}). Let $E\in\rr$ be such that $H^{-1}([E-\delta E, E+\delta E])$ is compact for some $\delta E>0$, and that
$\nrjt=\{ \widetilde{H}=E\}$ has no critical points.
We suppose that hypotheses of reduction are satisfied on $\ome\cap U$, where $U:=H^{-1}(]E-\delta E, E+\delta E[)$.
Let $f$ and $\psi$ be real functions in $\mathcal{S}(\rr)$ with $\supp(\psi)\subset ]E-\delta E,E+\delta E[$
and such that $\hat{f}$ is compactly supported.\\
If $\supp\hat{f}\cap\mathcal{L}_{red}(E)=\emptyset$, then  $\mathcal{G}_{\chi}(h)=O(h^{+\infty})$
as $h\to 0^+$.\\
If $\supp\hat{f}\cap\mathcal{L}_{red}(E)=\{ 0 \}$, then $\mathcal{G}_{\chi}(h)$ has a complete expansion
in powers of $h$, whose coefficients are distributions in $\hat{f}$ with support in $\{ 0 \}$, and:
\beq
\mathcal{G}_{\chi}(h)=(2\pi h)^{k_0-d+1}\hat{f}(0) \psi(E) \frac{d_{\chi}}{2\pi} 
\int_{\widetilde{\nrj}}dL_{\widetilde{H},E}\left[{\rho_{\chi}}_{|_{H_0}}:\indic\right]
+ O(h^{k_0-d+2}).
\eeq
where $dL_{\widetilde{H},E}=\frac{d\widetilde{\nrj}}{d \widetilde{H}}$ is the Liouville measure
associated to $\widetilde{H}$ on $\nrjt$, $k_0$ is the common dimension of $G$-orbits of $\ome\cap U$, and
$\left[{\rho_{\chi}}_{|_{H_0}}:\indic\right]$ is as in Theorem \ref{weak}.
\end{theo}
Namely, if $f$ is smooth with compact support in $(\ome\cap U)/G\subset \omer$, then the Liouville measure
satisfies: 
\beq\label{Liouville}
\int_{\widetilde{\nrj}} f(x) dL_{\widetilde{H},E}=\int_{\nrj\cap \ome}\frac{f(\pi(z))}{Vol(G(z))}
\frac{d\sigma_{\nrj\cap\ome}(z)}{\norm{\Pi_{T_z\ome}(\grad)}}_{\rdd}.
\eeq
where $d\sigma_{\nrj\cap\ome}$ is the Lebesgue measure on $\nrj\cap\ome$, and $\Pi_{T_z\ome}$ denotes the
orthogonal projector on $T_z\ome$ in $\rdd$.\\\\
{\it Remark:} as a consequence of this theorem, we note that if $\hat{f}$ is supported close enough to
zero, then $\supp\hat{f}\cap\mathcal{L}_{red}(E)=\{ 0 \}$ (because of the existence of a minimal period
on $\nrjt$). Using a classical Tauberian argument (see \cite{Ro}), this leads to Theorem
\ref{Comptage}.\\\\
\underline{{\it Proof:}} the case where $\supp\hat{f}\cap\mathcal{L}_{red}(E)=\emptyset$ is straightforward
by using a non-stationary phase theorem, since the intersection of the critical set with the support of the
amplitude is empty.
Suppose that $\supp\hat{f}\cap\mathcal{L}_{red}(E)=\{ 0 \}$. Let $(0,z,g)\in\ce \cap
(\supp(\hat{f})\times\rdd\times G)$. Let $(\tau,\aaa,Ag)\in \ker_{_{\rr}} \mbox{Hess} \, \vfi_E (0,z,g)$, i.e., by
Proposition \ref{calculnoyau}, $\aaa\perp \grad$, $\aaa\in (J\alg z)^{\perp}=T_z \ome$ and 
\beq\label{manodou}
\tau\jgrad+(\mg-I) \aaa +Az=0,
\eeq
Since $\mg z=z$, we have $(\mg-I)\aaa\in\alg z$. Thus, if $\tau \neq 0$, then $\jgrad\in\alg z$, which is in
contradiction with Lemma \ref{transversecrit}. Thus $\tau=0$. Therefore, in view of Proposition
\ref{cleanzero}, we have $\ker_{_{\rr}} \mbox{Hess} \, \vfi_E (0,z,g)=T_{(0,z,g)}\ce$, and we can apply the
stationary phase theorem.\

Now, we wish to compute the leading term of the expansion of $\mathcal{G}_{\chi}(h)$. We are going to use weak
asymptotics given in Theorem \ref{weak}. Note that we could have made the calculus using the result
of the stationary phase theorem, but the computation is more technical (see \cite{these}).\

Since $\mathcal{G}_{\chi,h}(E):=Tr\left(\psi(\hq_{\chi}) f\left( \frac{E-\hq_{\chi}}{h} \right)
\right)$ is continuous with respect to variable $E$, and since hypotheses of Theorem \ref{Weyl} are available for
energies in a neighbourhood of $E$, we have a continuous function $\la \mapsto a(\la)$ in the neighbourhood
of $E$ such that:
$$\mathcal{G}_{\chi,h}(\la)=a(\la)\, h^{k_0-d+1} + O(h^{k_0-d+2}),$$
with uniform rest with respect to $\la$ near $E$. Let $\vfi$ be a smooth compactly supported function  in this
neighbourhood. We have:
\beq\label{joli}
\int_{\rr} \vfi(\la)\mathcal{G}_{\chi,h}(\la)d\la=h^{k_0-d+1}\int_{\rr} \vfi(\la)a(\la)d\la + O(h^{k_0-d+2}).
\eeq
Moreover, if $x\in\rr$, then $\int_\rr \vfi(\la) f\left(\frac{\la-x}{h}\right)d\la=
h\int_\rr\vfi(x+t h) f(t)dt.$
Thus:
\beq\label{tele}
\int_{\rr} \vfi(\la)\mathcal{G}_{\chi,h}(\la)d\la=
h\int_\rr Tr\left( \psi(\hq_{\chi})\vfi(\hqr+t h)\right)f(t)dt.
\eeq
If $x$ and $\la$ are real, we have $|\vfi(x+t h)-\vfi(x)|\leq \norm{\vfi^{\prime}}_{\infty}|t h|.$
Therefore:
\begin{tabbing}
$\left|Tr\left( \psi(\hq_{\chi})[\vfi(\hqr+t h)-\vfi(\hqr)]\right)\right|$
\= $\leq \norm{\psi(\hq_{\chi})}_{Tr} \norm{\vfi(\hqr+t h)-\vfi(\hqr)}_{\mathcal{L}(\lde)}$\\
\> $\leq\norm{\psi(\hq_{\chi})}_{Tr}\norm{\vfi^{\prime}}_{\infty}|t h|$
\end{tabbing}
$\psi$ being non negative, we have $\norm{\psi(\hq_{\chi})}_{Tr}=Tr(\psi(\hq_{\chi}))=O(h^{k_0-d})$ in view of
the weak asymptotics. Thus
$$\int_\rr Tr\left( \psi(\hq_{\chi})[\vfi(\hqr+t h)-\vfi(\hqr)]\right)dt =O(h^{k_0-d+1}).$$
Using (\ref{tele}), we get:
$$ \int_{\rr} \vfi(\la)\mathcal{G}_{\chi,h}(\la)d\la
=h\hat{f}(0)Tr\left( \psi(\hq_{\chi})\vfi(\hqr)\right)+O(h^{k_0-d+2}).$$
Thus, by Theorem \ref{weak}, we have:
$$\int_{\rr}
\vfi(\la)\mathcal{G}_{\chi,h}(\la)d\la=h^{k_0-d+1}(2\pi)^{k_0-d}d_{\chi}\hat{f}(0)\int_{\omer}(\psi\vfi)(\tilde{H}(x))
d\sigma_{red}(x)\left[{\rho_{\chi}}_{|_{H_0}}:\indic\right] + O(h^{k_0-d+2})$$
which implies that (in view of (\ref{joli})):
$$\int_{\rr} \vfi(\la)a(\la)d\la=(2\pi)^{k_0-d}d_{\chi}\hat{f}(0)\int_{\omer}(\psi\vfi)(\tilde{H}(x))
d\sigma_{red}(x)\left[{\rho_{\chi}}_{|_{H_0}}:\indic\right].$$
Thus, by (\ref{intomer}):
\beq\label{mathou}
\int_{\rr}
\vfi(\la)a(\la)d\la=(2\pi)^{k_0-d}d_{\chi}\hat{f}(0)\int_{\ome}\frac{(\psi\vfi)(H(z))}{Vol(G(z))}
d\sigma_{\ome}(x)\left[{\rho_{\chi}}_{|_{H_0}}:\indic\right].
\eeq
Using Lemma \ref{lemmetranche}, one can show that, if $f$ is smooth, compactly supported in $\ome\cap U$,
then:
$$\int_{\ome} f(z) d\sigma_{\ome}(z)=\int_\rr \int_{\ome\cap \Sigma_\la} f(z)
\frac{d\sigma_{\ome\cap \Sigma_\la}(z)}{\norm{\Pi_{T_z\ome}(\grad)}_{\rdd}} d\la,$$
where $\Pi_{T_z\ome}$ denotes the orthogonal projector on $T_z\ome$.
Finally, one can apply this last formula to (\ref{mathou}) to get:
$$\int_{\rr} \vfi(\la)a(\la)d\la=(2\pi)^{k_0-d}d_{\chi}\hat{f}(0)\left[{\rho_{\chi}}_{|_{H_0}}:\indic\right]\int_\rr\vfi(\la)\psi(\la)
\int_{\ome}  \frac{1}{Vol(G(z))}\frac{d\sigma_{\ome\cap \Sigma_\la}(z)}{\norm{\Pi_{T_z\ome}(\grad)}_{\rdd}} d\la,$$
and we get the expression of $a(E)$.\qed
\section{The oscillatory part}\label{Gutzpart}
This section is dedicated to the proof of Theorem \ref{Gutz}. We suppose that $0\notin \supp(\hat{f})$ and
that periodic orbits of $\nrjt$
having a period in $\supp(\hat{f})$ are {\it non degenerate}, i.e., if $x\in\nrjt$ and $\tilde{T}\in\supp(\hat{f})$
satisfy $\tilde{\Phi}_{\tilde{T}}(x)=x$, then $1$ is not an eigenvalue of the differential of the Poincar\'e map at $x$
at time $\tilde{T}$ restricted to the energy level $\nrjt$. If $x\in\nrjt$, we denote by $\widetilde{F}_x(\tilde{T})$
the differential of the reduced flow $x\mapsto \widetilde{\Phi}_{\tilde{T}}(x)$ with respect to
variable $x\in\omer$. The non degenerate hypothesis is equivalent to say that, if
$\tilde{\Phi}_{\tilde{T}}(x)=x$, with $\tilde{T}\in \supp(\hat{f})$ then:
\beq\label{nondegenerate1}
\dim\ker[(\widetilde{F}_x(\tilde{T})-Id_{T_x\omer})^2]\leq 2.
\eeq
Note that this is always the case when $\dim(\omer)=2$, i.e. $d=k_0+1$. In particular this happens for
the spherical symmetry $G=SO(d)$ (see examples in section 2).
\subsection{The stationary phase problem}
Since $\nrjt$ is compact without critical points, then by the cylinder theorem (see \cite{Ab-Ma} 8.2.2), there
is a finite number of periodic orbits in $\nrjt$ with period in $\supp(\hat{f})$ and $\mathcal{L}_{red}\cap
\supp(\hat{f})$ is a finite set. If $t_0\in\mathcal{L}_{red}\cap \supp(\hat{f})$, we set:
\beq
W_{t_0}:=\{ (z,g)\in (\ome\cap\nrj)\times G \; :  \; \mg\Phi_{t_0}(z)=z  \}.
\eeq
By Proposition \ref{calculenscrit}, we get that:
\beq\label{ceND}
\ce\cap (\supp(\hat{f})\times\rdd\times G)=
\bigcup_{t_0\in\mathcal{L}_{red}\cap supp(\hat{f})} \{ t_0 \} \times W_{t_0}
\eeq
\begin{prop}\label{Wt0}
We suppose that periodic orbits of $\nrjt$ are non degenerate on $\supp(\hat{f})$ in the sense given above.
Let $t_0 \in\mathcal{L}_{red}\cap \supp(\hat{f})$.
Then $W_{t_0}$ is a submanifold of $\rdd\times G$, $\dim W_{t_0}=p+1$, and if $(z,g)\in W_{t_0}$, then:
\beq\label{tangentWt0}
T_{(z,g)}W_{t_0}=\{(\aaa,Ag)\in T_z\ome\times\alg g : \aaa\in\rr \jgrad +\alg z\mbox{ and } 
(\mg F_z(t_0)-I)\aaa+Az=0\}.
\eeq
\end{prop}
\underline{{\it Proof:}} let $\pi_1:\ome\cap \nrj\to\nrjt$ be the restriction of $\pi$ to $\nrj\cap \ome$. We
claim that $\pi_1$ is a submersion. Indeed, if $z\in\ome\cap \nrj$, and if $u\in T_{\pi(z)}\nrjt$, then, $\pi$
being a submersion, there exists $\aaa\in T_z\ome$ such that $u=d_z\pi(\aaa)$.
By the Theorem \ref{symplecticred}, we have: 
$$d_{\pi(z)}\tilde{H}(d_z\pi(\aaa))=\omega_{red}(\pi(z))(d_z\pi(\aaa),X_{\tilde{H}}(\pi(z)))
=<J\aaa,\jgrad>,$$
since $d_z\pi(\jgrad)=X_{\tilde{H}}(\pi(z))$. Remembering that $T_{\pi(z)}\nrjt=\ker d_{\pi(z)}\tilde{H}$, we
get that $\aaa\perp\grad$, and $\aaa\in (T_z\ome)\cap(T_z\nrj)=T_z(\ome\cap\nrj)$.\

If $z\in\ome\cap U$, if $g\in G$, differentiating with respect to $z\in\ome$ the identity $\pi(\mg
\Phi_{t_0}(z))=\tilde{\Phi}_{t_0}(\pi(z))$, we get that, if $z=\mg\Phi_{t_0}(z)$,
then $\mg F_z(t_0) \, T_z\ome\subset T_z\ome$, and we have on $T_z\ome$:
\beq\label{monored}
d_z\pi \circ \mg F_z(t_0)=\widetilde{F}_{\pi(z)}(t_0)\circ d_z\pi.
\eeq
Let $R_{t_0}:=\{ (z,\mg\Phi_{t_0}(z)):z\in\nrj\cap\ome,g\in G \}$.
\begin{lem}\label{Rto}
$R_{t_0}$ is a submanifold of $(\nrj\cap\ome)^2$, $\dim
R_{t_0}=2d-1$. If $z=\mg\Phi_{t_0}(z)$ then
$$T_{(z,z)}R_{t_0}=\{(\aaa,\mg F_{z}(t_0)\aaa+Az):\aaa\in
T_z(\ome\cap\nrj) \mbox{ and } A\in \alg \}.$$
\end{lem}
\underline{{\it Proof:}} let $\A_{t_0}:=\{ (x,\tilde{\Phi}_{t_0}(x)) :x\in\nrjt \}.$
$\A_{t_0}$ is a submanifold of $[(\ome\cap U)/G]^2$ and $\dim
\A_{t_0}=\dim\nrjt=2d-2k_0-1$. Moreover $R_{t_0}=(\pi_1\times\pi_1)^{-1}(\A_{t_0})$.
Thus, $\pi_1\times\pi_1$ being submersion, $R_{t_0}$ is a submanifold $(\ome\cap U)^2$ and $\dim R_{t_0}=2d-1$.
Besides, $T_{(x,\tilde{\Phi}_{t_0}(x))}\A_{t_0}=\{(u,\widetilde{F}_x(t_0)u) :
u\in T_x\nrjt \}$. Thus, if $z=\mg \Phi_{t_0}(z)$, then:
$$T_{(z,z)}R_{t_0}=\{(\aaa,\beta) \in T_{(z,z)}(\ome\cap\nrj)^2 :
d_z\pi(\beta)= \widetilde{F}_{\pi(z)}(t_0)d_z\pi(\aaa)\}.$$
In view of (\ref{monored}),
the proof is clear if we remember that $\ker d_z\pi=\alg z$.\qed\\\

We denote by $\mathcal{P}_{red}(E,t_0)$ the set of periodic orbits of $\nrjt$ with period $t_0$. We have:
\beq\label{orbitperiodicwto}
W_{t_0}=\underset{\bar{\gmm}\in\mathcal{P}_{red}(E,t_0)}{\bigsqcup} \Lambda_{\bar{\gmm},t_0}.
\eeq
where
\beq\label{lambdabar1}
\Lambda_{\bar{\gmm},t_0}:=\{ (z,g)\in (\ome\cap\nrj)\times G \; :  \; z=\mg\Phi_{t_0}(z) \;\mbox{ and }\;
\pi(z)\in\bar{\gmm}  \}.
\eeq
Let $\bar{\gamma}\in \mathcal{P}_{red}(E,t_0)$ and $\vfi_0:(\ome\cap\nrj)\times G \to R_{t_0}$ defined by
$\vfi_0(z,g):= (z,\mg\Phi_{t_0}(z))$. We have $\Lambda_{\bar{\gmm}}=\vfi_0^{-1}(P_{\bar{\gamma}})$, where
$P_{\bar{\gamma}}:= \{(z,z)\in(\nrj\cap \ome)^2 : \pi(z)\in\bar{\gamma} \} \subset R_{t_0}.$ We have:
$P_{\bar{\gamma}}=diag(\pi_1^{-1}(\bar{\gamma}))$, thus, $\pi_1$ being a submersion,
$P_{\bar{\gamma}}$ is a submanifold of $R_{t_0}$, $\dim(P_{\bar{\gamma}})=k_0+1$, and
$T_{(z,z)}P_{\bar{\gamma}}=(\alg z+\rr\jgrad)^2$, since we have $d_z\pi(\jgrad)=X_{\tilde{H}}(\pi(z))$ and 
$\ker d_z\pi=\alg z$. The proof of the proposition is clear if we show that $\vfi_0$ is a submersion at all
points of $W_{t_0}$, which is given by Lemma \ref{Rto}.\qed\\\

Now we have to show that the transversal Hessian of $\vfi_E$ is non degenerate on points of $\{t_0\}\times W_{t_0}$.
Let $(z,g)\in W_{t_0}$. We have to show that $\ker_{_{\rr}}\mbox{Hess}\; \vfi_E(t_0,z,g)=\{ 0 \}\times
T_{(z,g)} W_{t_0}$. Let $\tau\in\rr$, $\aaa\in T_z(\ome\cap\nrj)$, and $A\in\alg$ such that:
\beq\label{crocodil}
\tau\jgrad+(\mg F_z(t_0)-Id) \aaa+Az=0.
\eeq
Applying $d_z\pi$, we get (using (\ref{monored})):
\beq\label{youpala}
\tau X_{\tilde{H}}(\pi(z)) + (\widetilde{F}_{\pi(z)}(t_0)-Id)
     d_z\pi (\aaa)=0.
\eeq
Thus, $d_z\pi (\aaa)\in \ker(\widetilde{F}_{\pi(z)}(t_0)-Id_{\omer})^2$. Using by (\ref{nondegenerate1}), we
have a basis $(u_1,u_2)$ of $\ker(\widetilde{F}_{\pi(z)}(t_0)-Id_{\omer})^2$ with:
$u_1=X_{\tilde{H}}(\pi(z)) \mbox{ et } w_{red}(\pi(z)) (u_1,u_2)\neq 0$ (as in the case without symmetry).
Thus, there exists $\la_1,\la_2$ in $\rr$ such that $d_z\pi(\aaa)=\la_1 u_1+\la_2 u_2$. Therefore:
$$w_{red}(\pi(z))(d_z\pi(\aaa),u_1)=d_{\pi(z)}\tilde{H}(d_z\pi(\aaa))=<\grad,\aaa>=0.$$
Thus $\la_2=0$ and $d_z\pi(\aaa)=\la_1 u_1$. Using (\ref{youpala}), we obtain $\tau=0$ since there is no critical
points of $\widetilde{H}$ on $\nrjt$. Besides, since $d_z\pi(\jgrad)=X_{\tilde{H}}(\pi(z))$, we have
$\aaa\in\rr\jgrad+\alg z$, and, in view of (\ref{crocodil}) and Proposition \ref{Wt0}, we get that $(0,\aaa,Ag)\in T_{(t_0,z,g)}\ce$.\\\

Thus, we have shown that, when periodic orbits of the reduced space are non degenerate, then we can apply the
stationary phase theorem, and get an asymptotic expansion of $\mathcal{G}_\chi (h)$.
\subsection{Computation of first terms}
We apply the stationary phase formula to (\ref{firstterm}). The set
$\ce\cap (\supp(\hat{f})\times\rdd\times G)$ splits into disjoint sets 
$\{t_0\}\times \Lambda_{\bar{\gmm},t_0}$ given by (\ref{ceND}) and (\ref{orbitperiodicwto}), where
$\Lambda_{\bar{\gmm},t_0}$ is  given by (\ref{lambdabar1}). $\Lambda_{\bar{\gmm},t_0}$ may not be a connected set, but
one can show that the quantity $S(t_0)(z):=\int_0^{t_0}pdq$ doesn't depend on $z\in\pi^{-1}(\bar{\gmm})$ (see
\cite{Ar}). We denote it by $S_{\bar{\gmm}}(t_0)$.
Then, using that $\dim \Lambda_{\bar{\gmm},t_0}=\dim W_{t_0}=p+1$, we get that:
$$\mathcal{G}_{\chi}(h)=\psi(E)d_{\chi}
\sum_{{\scriptsize t_0\in\mathcal{L}_{red}(E)\cap\supp\hat{f} }} \hat{f}(t_0)
\sum_{\bar{\gmm}\in \mathcal{P}_{red}(E,t_0)} e^{\frac{i}{h}S_{\bar{\gmm}}(t_0)} 
\frac{1}{2\pi}
\int_{\Lambda_{\bar{\gmm},t_0}} \carb d(t_0,z,g) d\sigma_{\Lambda_{\bar{\gmm}}}(z,g)
+ O(h),$$
where $d(t_0,z,g)$ is given by:
$$d(t,z,g):= \detp\left( \frac{\varphi_{E}^{\prime\prime}(t,z,g)_{|_{\mathcal{N}_{(t,z,g)}\ce}}}{i} \right)
\detp\left( \frac{A+iB-i(C+iD)}{2}\right).$$
In this case, the computation of $\detp\left(
\frac{\varphi_{E}^{\prime\prime}(t,z,g)_{|_{\mathcal{N}_{(t,z,g)} \ce}}}{i} \right)$ seems to be a non trivial
calculus. We didn't succeed in interpreting it geometrically as in the case of the Weyl term. However, one
should be able to make appear the primitive period of $\bar{\gamma}$ and the differential of its Poincar\'e
map, as we did in the case of a finite group in \cite{fini}.


\begin{thebibliography}{99}
 
\bibitem{Ab-Ma} R. Abraham, J.E. Marsden, \emph{Foundations of mechanics}, The
Benjamin/Cummings Publishing Company, Inc (1978).

\bibitem{Ar} V.I. Arnold, \emph{Mathematical methods of classical mechanics},
Graduate Texts in Mathematics. 60. New York - Heidelberg - Berlin: Springer-Verlag.

\bibitem{B-P-U} D. Borthwick,  T. Paul, A. Uribe, \emph{Semiclassical spectral estimates for Toeplitz
operators}, Ann. Inst. Fourier 48, No.4, 1189-1229 (1998). 

\bibitem{B-H} J. Br\"uning, E. Heintze, \emph{Representations of compact Lie groups and elliptic
operators}, Invent. Math. {\bf 50}, 169-203 (1979).

\bibitem{B-Z} Y.D. Burago, V.A. Zalgaller, \emph{Geometric inequalities}, Springer Series in Soviet
Mathematics. Springer-Verlag, Berlin (1988).

\bibitem{cras1} R. Cassanas, \emph{A Gutzwiller type formula for a reduced Hamiltonian within the framework
of symmetry} C. R., Math., Acad. Sci. Paris 340, No.1, 21-26 (2005).


 \bibitem{fini} R. Cassanas, \emph{Reduced Gutzwiller formula with symmetry: case of a finite group},
submitted.

\bibitem{these} R. Cassanas, \emph{Hamiltoniens quantiques et sym\'etries}, PhD Thesis, Universit\'e de
Nantes, (2005). Available on the web site: http://tel.ccsd.cnrs.fr/documents/archives0/00/00/92/89/index\_fr.html

\bibitem{Ch} L. Charles, \emph{Toeplitz operators and Hamiltonian Torus Action}, Preprint, 2004.

\bibitem{CRR} M. Combescure, J. Ralston, D. Robert, \emph{A proof of the Gutzwiller semiclassical
trace formula using coherent states decomposition}, Commun. Math. Phys., {\bf 202}, 463-480, (1999).

\bibitem{CR} M. Combescure, D. Robert, \emph{Semiclassical spreading of quantum wave packets and
applications near unstable fixed point of the classical flow}, Asymptotic Anal. {\bf 14},377-404, (1997).

\bibitem{Cr} S.C. Creagh, \emph{Semiclassical mechanics of symmetry reduction}, J. Phys. A, Math.
Gen. 26, No.1, 95-118 (1993).

\bibitem{Don} H. Donnelly: \emph{G-spaces, the asymptotic splitting of $L^2(M)$ into irreducibles},
Math. Ann. {\bf 237}, pp.23-40, (1978).

\bibitem{El.H} Z. El Houakmi, \emph{Comportement semi-classique du spectre en pr\'esence de
sym\'etries : Cas d'un groupe fini}, Th\`ese de 3\`eme cycle et S\'eminaire de Nantes (1984).

\bibitem{El.H-He} Z. El Houakmi, B. Helffer, \emph{Comportement semi-classique en pr\'esence de
sym\'etries. Action d'un groupe compact}, Asymptotic Anal. {\bf 5}, No.2, 91-113 (1991).
 
\bibitem{G-U 1} V. Guillemin, A. Uribe, \emph{Reduction, the trace formula, and semiclassical asymptotics}
Proc. Natl. Acad. Sci. USA {\bf 84}, 7799-7801 (1987).

\bibitem{G-U 2} V. Guillemin, A. Uribe, \emph{Circular symmetry and the trace formula}, Invent. Math. 96,
No.2, 385-423 (1989).

\bibitem{G-U 3} V. Guillemin, A. Uribe, \emph{Reduction and the trace formula}, J. Differ. Geom. 32, No.2,
315-347, (1990).

\bibitem{He-Ro 1} B. Helffer et D. Robert, \emph{Calcul fonctionnel par la tansform\'ee de Mellin},
 J. Funct. Anal, {\bf 53}, 246-268 (1983).

\bibitem{He-Ro 2} B. Helffer et D. Robert, \emph{Etude du spectre pour un \ope globalement
elliptique dont le symbole de Weyl pr\'esente des sym\'etries I: Action des groupes finis.},
Am. J. Math. {\bf 106}, 1199-1236 (1984).

\bibitem{He-Ro 3} B. Helffer et D. Robert, \emph{Etude du spectre pour un \ope globalement
elliptique dont le symbole de Weyl pr\'esente des sym\'etries II: Action des groupes de Lie compacts.},
Amer. J. of Math., {\bf 108}, 973-1000 (1986).

\bibitem{K} K. Kawakubo (1991) : \emph{The theory of transformation groups}, Oxford University Press, Oxford, UK.

\bibitem{Or-Ra} J.P. Ortega, T.S. Ratiu, \emph{Momentum maps and Hamiltonian reduction}, Progress
in Math., vol.222, Bikh\"{a}user (Boston, Mass.) (2004).

\bibitem{Pi} G. Pichon, \emph{Groupes de Lie. Repr\'esentations lin\'eaires et applications},
Hermann, Paris (1973).

\bibitem{Ro} D. Robert, \emph{Autour de l'approximation semi-classique}, Progress in Math.,
vol.{\bf 68}, Bikh\"{a}user, Basel (1987).

\bibitem{Ro1} D. Robert, \emph{Remarks on Asymptotic solutions for time dependent Schr\"odinger
equations}, Optimal Control and Partial Differential Equations, IOS Press p.188-197 (2001).

\bibitem{Se} J.P. Serre, \emph{Repr\'esentations lin\'eaires de groupes finis}, Hermann, Paris (1967).

\bibitem{Si} B. Simon, \emph{Representations of finite and compact groups}, Graduate Studies
in Math., Amer. Math. Soc. (1996).


\end{thebibliography}
\end{document}